\documentclass[prd,preprintnumbers,showkeys,nofootinbib, 11pt]{article}
\usepackage{graphicx,amsmath,amssymb,epsf}
\usepackage{latexsym,tabularx,bm, slashed}
\newcommand{\tr}{{\rm tr}} 

\begin{document}

\title{ \bf Quark contribution to the gluon Regge trajectory at NLO from the high energy effective action\footnote{Preprint numbers: LPN12-014, IFT-UAM/CSIC-12-03, FTUAM-12-81}}

\author{G.~Chachamis$^1$, M.~Hentschinski$^2$, J.~D.~Madrigal Mart{\' \i}nez$^2$, A.~Sabio Vera$^2$ 
\bigskip \\
{ $^1$~Paul Scherrer Institut, CH-5232 Villigen PSI, Switzerland.}\\
{ $^2$~Depto. de F{\' 	\i}sica Te{\' o}rica \& Instituto de F{\' \i}sica Te{\' o}rica UAM/CSIC,}\\
{Universidad Aut{\' o}noma de Madrid,  Cantoblanco E-28049 Madrid, Spain.}
}
%\date{}
\maketitle

\begin{abstract}
The two loop (NLO) diagrams with quark content contributing to the gluon Regge trajectory are computed within the framework of Lipatov's effective action for QCD, using the regularization procedure for longitudinal
  divergencies recently proposed by two of us in~\cite{Hentschinski:2011tz}. 
  Perfect agreement with previous results in the literature is 
  found, providing a robust check of the regularization prescription and showing that the high energy effective
  action is a very useful computational tool in the quasi-multi-Regge limit.
\end{abstract}

\section{Introduction}
\label{1}

Very useful all-orders information can be obtained for elastic scattering amplitudes in Quantum Chromodynamics (QCD) when the interaction takes place in the high energy Regge limit of large center-of-mass energy. For inelastic processes a generalization is to consider multi-Regge kinematics where the Regge limit is applied to multiple sub-channels~\cite{BFKL1}. As it is well 
known, in this regime new degrees of freedom emerge which correspond to reggeized gluons exchanged in the 
$t$-channel. Their propagators get modified by a factor of the form $s_i^{\omega(t_i)}$ indicating the no-emission probability of any particle in the rapidity interval $\sim \log(s_i/|t_i|)$, where each sub-channel has an invariant mass $\sqrt{s_i}$ and momentum transfer $t_i$. The function controlling this no-emission probability is the 
reggeized gluon trajectory $\omega(t_i)$ which has been calculated at leading ${\cal O} \left({\alpha_s}\right)$ (LO) and next-to-leading order ${\cal O} \left({\alpha_s^2}\right)$  (NLO) in QCD and to all orders in ${\cal N}=4$ super Yang-Mills theory~\cite{Trajectory}.  Emission probabilities are given by effective vertices for the interaction of reggeized gluons with usual particles. At NLO $\alpha_s \left(\alpha_s \log{(s_i/t_i)} \right)^n$ terms are resummed and in each production vertex a cluster with one or two particles is produced, generating the so-called quasi-multi-Regge kinematics (QMRK)~\cite{BFKLNLO}. 

At NLO scattering amplitudes can still be written in the form of a linear equation while at higher orders this 
linearity is broken since transition vertices with multiple reggeized gluons appear (see, {\it e.g.}~\cite{Bartels:1995kf}). 
The high energy effective action proposed by Lipatov~\cite{LevSeff} is a very efficient tool to evaluate these 
transition vertices. It is based on the QCD action together with an induced contribution written in terms of gauge-invariant currents where Wilson lines generate color fields ordered in longitudinal components, with non-trivial interactions with the fields representing the reggeized gluons. Some of the relevant vertices were calculated using the Feynman rules derived from this effective action (see~\cite{LevSeff} and~\cite{Martin}). 
In~\cite{Hentschinski:2011tz} two of us used this effective action to calculate the vertex for the transition of 
a quark into a forward jet plus a remnant together with an off-shell reggeon in the $t$-channel. This vertex is  important for applications of QMRK to hadron colliders~\cite{Vera:2006un}. 

Lipatov's effective action is not only useful to efficiently calculate transition and emission vertices but it can 
also be used to obtain the quark and gluon Regge 
trajectories. In this work we offer a calculation at two loops where 
the regularization of longitudinal divergencies presented in~\cite{Hentschinski:2011tz} allows us to 
obtain the quark contributions to the NLO gluon trajectory. This shows how the high energy action 
is indeed a very useful tool also to obtain complicated quantum corrections within this effective theory.  
The technically more involved cases of gluon contributions and the NLO quark Regge trajectory will be discussed in future publications. 

In this work a first section is given with a discussion on Lipatov's effective action together with the Feynman rules used in the subsequent calculations. Our main results are explained in Section 3 where our 
regularization and subtraction procedures are explained in detail, together with the final derivation of the 
gluon trajectory in the case of a given elastic amplitude. Finally, our conclusions and outlook for future work 
are presented.

\section{The effective action for QCD at high energy}
\label{sec:lagrangian}
In this Section we briefly introduce the notation used in our work and sketch the main features 
of Lipatov's effective action. 
For a general elastic scattering process at partonic level, with momenta $p_a +
p_b \to p_1 + p_2$ and $p_a^2 = p_b^2 = 0$, the squared center of mass energy is 
$s = (p_a + p_b)^2 =2 p_a\cdot p_b$. It is very useful to work with the 
light-like four-vectors $n^\pm$ fullfilling $n^+\cdot n^- = 2$ and being related to
the incoming momenta by the re-scaling $ n^+ = {2 p_b
}/{\sqrt{s}} $ and $ n^- = {2 p_a}/{\sqrt{s}} $. The Sudakov
decomposition of a general four-vector $k^\mu$ hence reads $ k = {k^+}
n^-/{2} + {k^-} n^+/{2} + {\bm k} $, where $k^\pm = n^\pm\cdot k $ and
${\bm k}$ is transverse to the initial scattering axis.  Compared to the usual QCD action, in Lipatov's approach one adds an induced term, $S_{\text{eff}} = S_{\text{QCD}} + S_{\text{ind}}$, which describes the coupling of a reggeized gluon field $A_\pm(x) = -i t^a A_\pm^a(x)$ to the usual gluon fields $v_\mu(x) = -it^a v_\mu^a(x)$, where $t^a$ are color matrices. This new component in the action reads
\begin{align}
\label{eq:1efflagrangian}
  S_{\text{ind.}}[v_\mu, A_\pm]& = \int \! d^4 x \,% \bigg\{
\tr\bigg[\bigg( W_+[v(x)] - A_+(x) \bigg)\partial^2_\perp A_-(x)\bigg]
\notag \\
&  \qquad  \qquad   \qquad
+\int \! d^4 x \, \tr\bigg[\bigg(W_-[v(x)] - A_-(x) \bigg)\partial^2_\perp A_+(x)\bigg]%\bigg\}
.
\end{align}
The infinite number of couplings of the usual to the reggeized gluon fields are implicitly written 
in the form of the two functionals $W_\pm[v] $, whose operator definition reads
$W_\pm[v] = v_\pm \frac{1}{ D_\pm}\partial_\pm$ where $D_\pm = \partial_\pm + g v_\pm $.

For perturbative calculations,  the following expansion in the gauge coupling $g$ holds, 
\begin{align}
  \label{eq:funct_expand}
  W_\pm[v] =&  v_\pm - g  v_\pm\frac{1}{\partial_\pm} v_\pm + g^2 v_\pm
\frac{1}{\partial_\pm} v_\pm\frac{1}{\partial_\pm} v_\pm - \ldots
\end{align}
One of the goals of this work is to provide a valid regularization of the $1/\partial_\pm$ operators 
which will allow to correctly calculate quantum corrections. It is important to note that the reggeized gluon fields are special in the sense that they are invariant under local gauge transformations, while they transform globally in the adjoint representation of the SU$(N_c)$ gauge group. In addition, the strong ordering of longitudinal momenta in the high energy amplitudes leads to the following kinematical constraint of the reggeized gluon fields,
\begin{align}
  \label{eq:constraint}
\partial_+ A_-(x)  =  \partial_- A_+(x) = 0,
\end{align} 
which is always implied. 

The quantization of the gluon field requires the addition to the QCD Lagrangian of a gauge fixing $\mathcal{L}_{\text{fix}}$ and a ghost term $\mathcal{L}_{\text{ghost}}$, 
which, for simplicity, we include in the QCD action, {\it i.e.}
\begin{align}
  \label{eq:action_QCD}%%%%% mmmma
{S}_{\text{QCD}} = 
 \int d^4 x  \big[\mathcal{L}_{\text{QCD}}& (v_\mu, \psi, \bar{\psi}) + \mathcal{L}_{\text{fix}}(v_\mu) + \mathcal{L}_{\text{ghost}}(v_\mu, \phi, \phi^\dagger)\big],
\end{align}
with $\phi$ being the ghost and $\psi$ the quark fields.

\begin{figure}[htb]
  \centering
   \parbox{.7cm}{\includegraphics[height = 1.8cm]{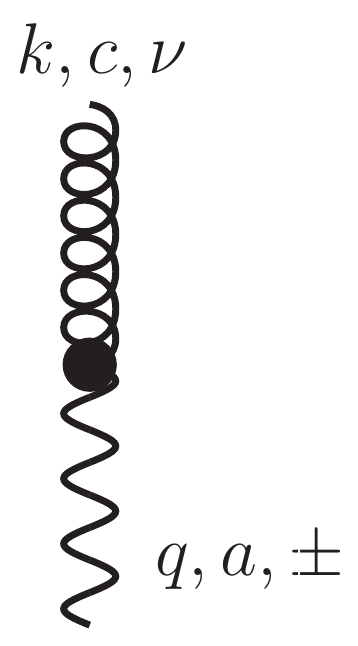}} $=  \displaystyle 
   \begin{array}[h]{ll}
    \\  \\ - i{\bm q}^2 \delta^{a c} (n^\pm)^\nu  \\ \\  \qquad   k^\pm = 0
   \end{array}  $ 
$~~~~$ \parbox{1.2cm}{ \includegraphics[height = 1.8cm]{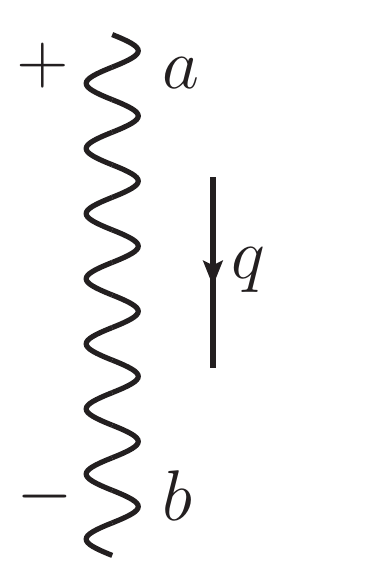}}  $=  \displaystyle    \begin{array}[h]{ll}
    \delta^{ab} \frac{ i/2}{{\bm q}^2} \end{array}$ 
$~~~~$ \ \parbox{1.7cm}{\includegraphics[height = 1.8cm]{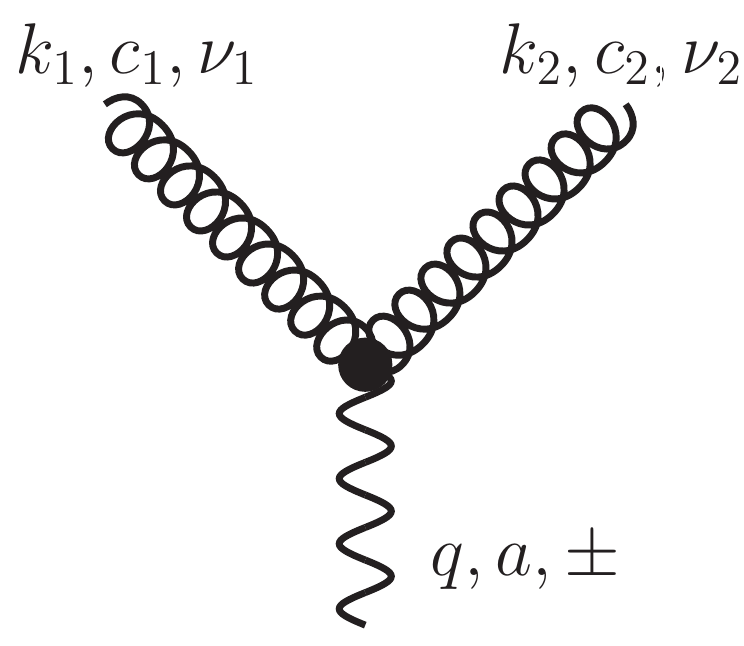}} $ \displaystyle  =  \begin{array}[h]{ll}  \\ \\ g f^{c_1 c_2 a} \frac{{\bm q}^2}{k_1^\pm}   (n^\pm)^{\nu_1} (n^\pm)^{\nu_2}  \\ \\ \quad  k_1^\pm  + k_2^\pm  = 0
 \end{array}$
 \\
\parbox{2cm}{\center (a)~~~~~~~~~~~~~~~} \parbox{4cm}{\center (b)} \parbox{5cm}{\center ~~~~~~~~~~(c)}
  \caption{\small (a) Direct transition vertex. (b) Reggeized gluon propagator.  (c) Unregulated  induced vertex at order $g$.}
  \label{fig:feynrules0p2}
\end{figure}

The Feynman rules for the high energy effective action have been derived 
in~\cite{LevSeff} and are given by the usual QCD Feynman rules 
together with an infinite number of induced vertices, 
which include a direct transition vertex from normal to reggeized gluons as it is shown in 
Fig.~\ref{fig:feynrules0p2} (a). We show them using curly 
lines for the conventional QCD gluon field and wavy lines for the reggeized gluon field.  
Using the direct transition vertex of Fig.~\ref{fig:feynrules0p2} (a) the propagator of the reggeized gluon receives a tree-level correction due to the projection of the gluon propagator
on the reggeized gluon states which is absorbed into the propagator of Fig.~\ref{fig:feynrules0p2} (b).  Due to the
expansion in Eq.~\eqref{eq:funct_expand} there exist an infinite number
of higher order induced vertices of which, for the present study, only the
order $g$ one in Fig.~\ref{fig:feynrules0p2} (c) is needed.  A possible tadpole  contribution from the 
${\cal O} (g^2)$ induced vertex to the reggeized gluon self energy vanishes due to the light-like gluon polarizations associated to the induced vertices.  For non light-like polarization vectors, the resulting loop integral is scaleless and vanishes within dimensional regularization. An explicit expression for the ${\cal O} (g^2)$ induced vertex  can be found, {\it e.g.}, in~\cite{Hentschinski:2011xg}.
In our evaluation of loop corrections it will be needed to fix a regularization of the light-cone singularity present in this vertex.  As suggested by one of us~\cite{Hentschinski:2011xg}, this pole 
will be treated as a Cauchy principal value in the following.

\section{The NLO gluon Regge trajectory: quark contribution}
\label{3}

As outlined in~\cite{Hentschinski:2011tz}, the evaluation 
of loop diagrams within the effective action approach leads to a new type of
longitudinal divergencies, which are not present in conventional quantum corrections to QCD
amplitudes. A convenient way to regularize these divergencies is to
introduce an external parameter $\rho$ which deforms the light-like four vectors
of the effective action in the form
\begin{eqnarray}
  \label{eq:n+-}
  n^- \to n_a &=& e^{-\rho} n^+ + n^- ,\\
  n^+ \to n_b &=&  n^+ + e^{-\rho}   n^-.
\end{eqnarray}
The deformation of the original reference momenta is then considered to be asymptotically 
small ($\rho \to \infty$) and $\rho$ can be given the interpretation of a logarithm of the 
center-of-mass energy. To regularize the remaining ultraviolet, soft and collinear divergencies,  
$d = 4 + 2 \epsilon$ dimensional regularization is used. Closely following the procedure 
proposed in~\cite{Hentschinski:2011tz} for the determination of the one loop quark contribution to the forward jet vertex, in the calculation of the two-loop gluon Regge trajectory we follow these steps:
\begin{itemize}
\item [ (i)] study of the one-loop reggeized gluon self-energy corrections for $\rho  \to \infty$
  up to ${\cal O} (\rho^0)$.
\item [ (ii)] study of the enhanced (${\cal O} (\rho, \rho^2)$ for $\rho  \to \infty$) contributions to the
  two-loop reggeized gluon self-energy.
\end{itemize}
In both cases the reggeized gluon is treated as a background field, 
{\it i.e.} no internal reggeized gluon propagators are included during 
the evaluation of these corrections. After these, the remaining tasks are
\begin{itemize}
\item [(iii)] subtraction of the non-local contribution from the two-loop
  result of (ii), which is given by two reggeized gluon self-energies joined 
  with a reggeized gluon propagator.
\item [(iv)] determination of the two-loop gluon Regge trajectory,
  from the high energy limit of a certain partonic two-loop QCD
  scattering amplitudes such as the quark-quark scattering, using the
  results (i) - (iii).
\end{itemize}
This procedure is general and in this work we focus on the evaluation of 
the quark contributions to the gluon trajectory 
to show its validity, leaving the more complicated gluon pieces for future work. 

\subsection{Results of the calculation and subtraction procedure}
\label{sec:results}

\begin{figure}[htb]
  \centering
  \parbox{1.5cm}{\vspace{0.1cm} \includegraphics[height = 2.5cm]{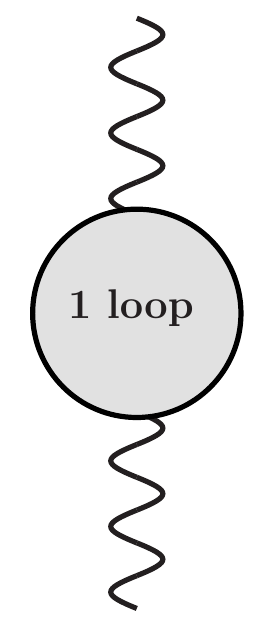}}
= 
    \parbox{1cm}{\vspace{0.1cm} \includegraphics[height = 2.5cm]{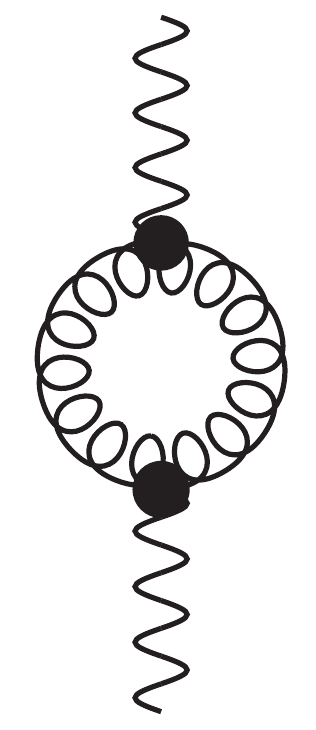}}
  + 
  \parbox{1cm}{\vspace{0.1cm} \includegraphics[height = 2.5cm]{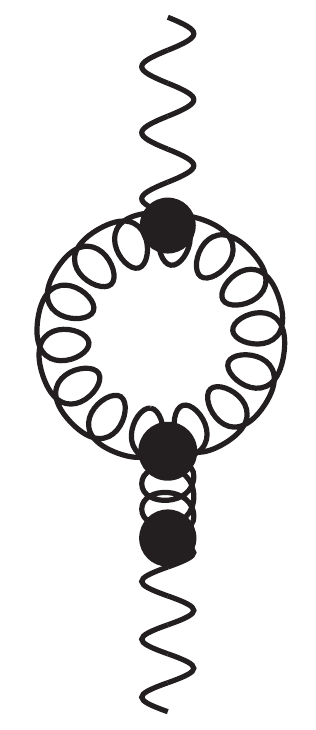}} 
 +
  \parbox{1cm}{\vspace{0.1cm} \includegraphics[height = 2.5cm]{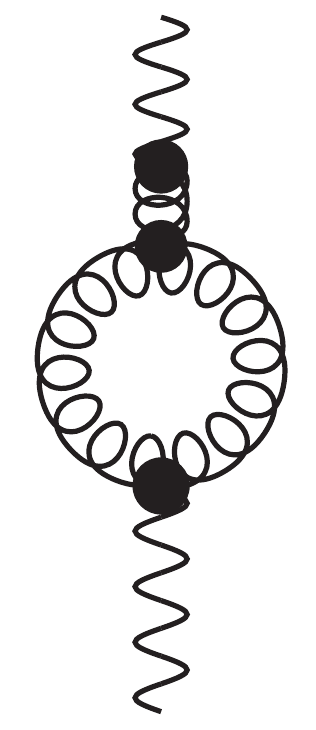}} 
 +
  \parbox{1cm}{\vspace{0.1cm} \includegraphics[height = 2.5cm]{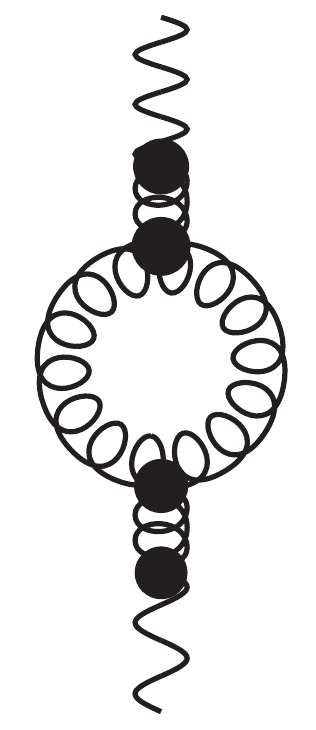}} 
 +
  \parbox{1cm}{\vspace{0.1cm} \includegraphics[height = 2.5cm]{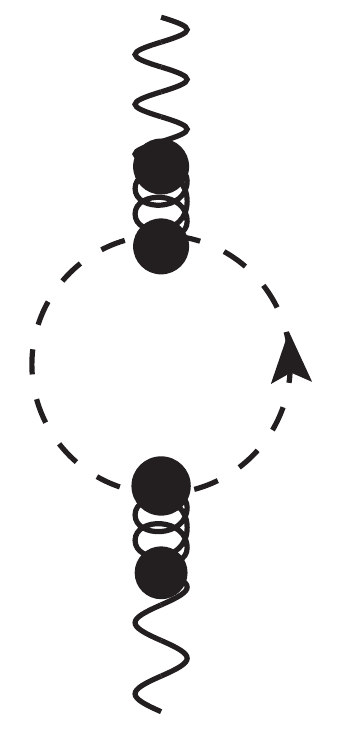}}
+
  \parbox{1cm}{\vspace{0.1cm} \includegraphics[height = 2.5cm]{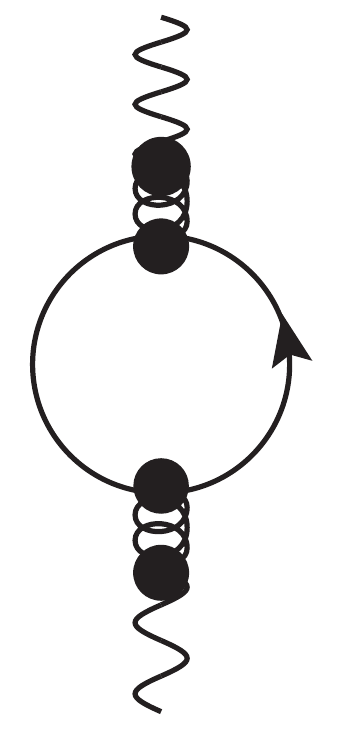}}
\caption{\small Diagrams contributing to the one-loop reggeized gluon self-energy.}
\label{fig:self_1loop}
\end{figure}

The one-loop reggeized gluon self-energy was presented in~\cite{Hentschinski:2011tz}. 
The contributing diagrams (including ghost loops) are shown in Fig.~\ref{fig:self_1loop}.
\\
Keeping the ${\cal O} (\rho, \rho^2)$, for $\rho  \to \infty$, terms and using the notation
\begin{align}
  \label{eq:gbar}
  \bar{g}^2 & =  \frac{g^2 N_c \Gamma(1 - \epsilon)}{(4 \pi)^{2 + \epsilon}},
\end{align}
we have the following result~\footnote{The conventions we follow for dimensional regularization  
can be found in the Appendix.}:
\begin{align}
\label{eq:self_1loop}
 &  \parbox{1cm}{\includegraphics[width = 1cm]{self_1loop.pdf}} %\hspace{-1.8cm} 
=    \Sigma^{(1)}\left(\rho; \epsilon, \frac{{\bm q}^2}{\mu^2}    \right)      % \hspace{-.3cm} 
\notag \\
& \hspace{.4cm}= 
 \frac{(-2i {\bm q}^2) \bar{g}^2 \Gamma^2(1 + \epsilon)}{\Gamma(1 + 2 \epsilon)} \left(\frac{{\bm q}^2}{\mu^2} \right)^\epsilon  
  \bigg\{  \frac{ i\pi - 2 \rho}{\epsilon}         
- \frac{1}{(1 + 2 \epsilon)\epsilon} \bigg[   2 
 +  \frac{5 + 3\epsilon}{3 + 2 \epsilon} 
-\frac{n_f}{N_c}  \left(\frac{2 + 2\epsilon}{3 + 2\epsilon}\right)\bigg] \bigg\}. 
\end{align}
As we mentioned above, in this work we are interested in the quark contributions at two loops. 
\begin{figure}
 % \centering
  \parbox{1.5cm}{\vspace{0.1cm} \includegraphics[height = 2.5cm]{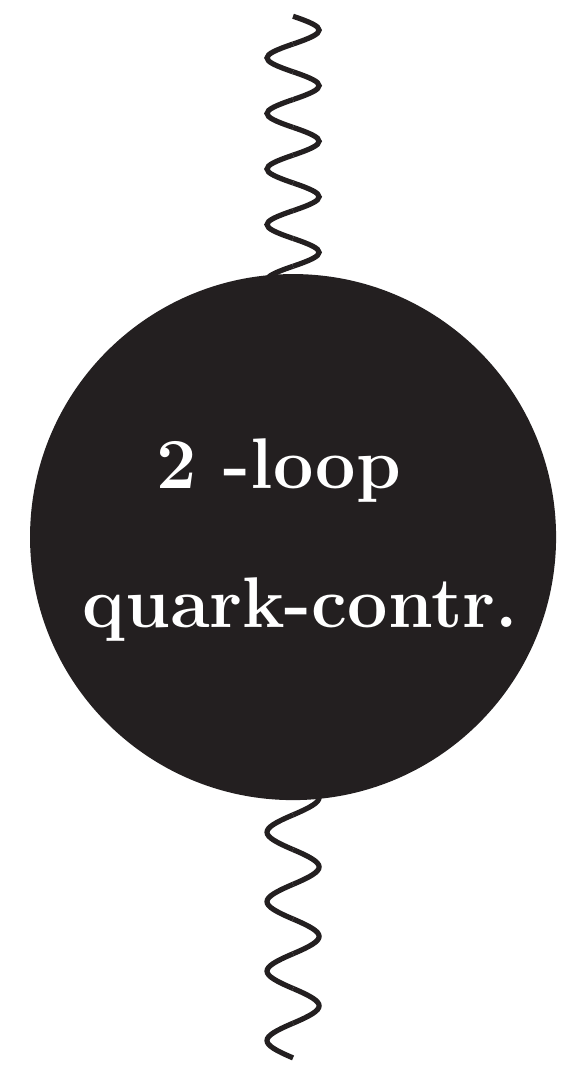}}
 = 
\parbox{1.4cm}{\vspace{0.1cm} \includegraphics[height = 2.5cm]{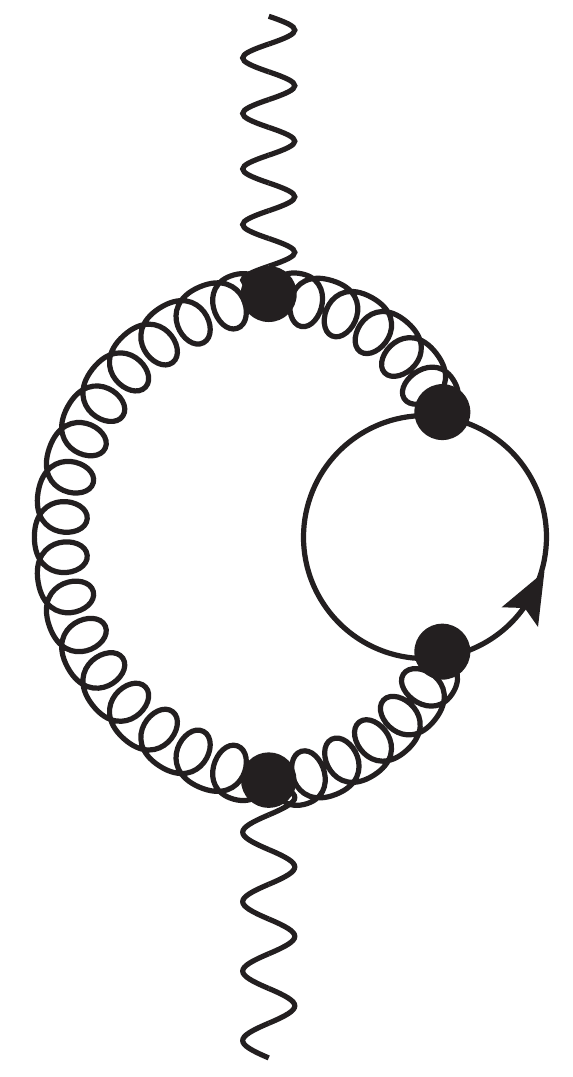}}
+
\parbox{1.4cm}{\vspace{0.1cm} \includegraphics[height = 2.5cm]{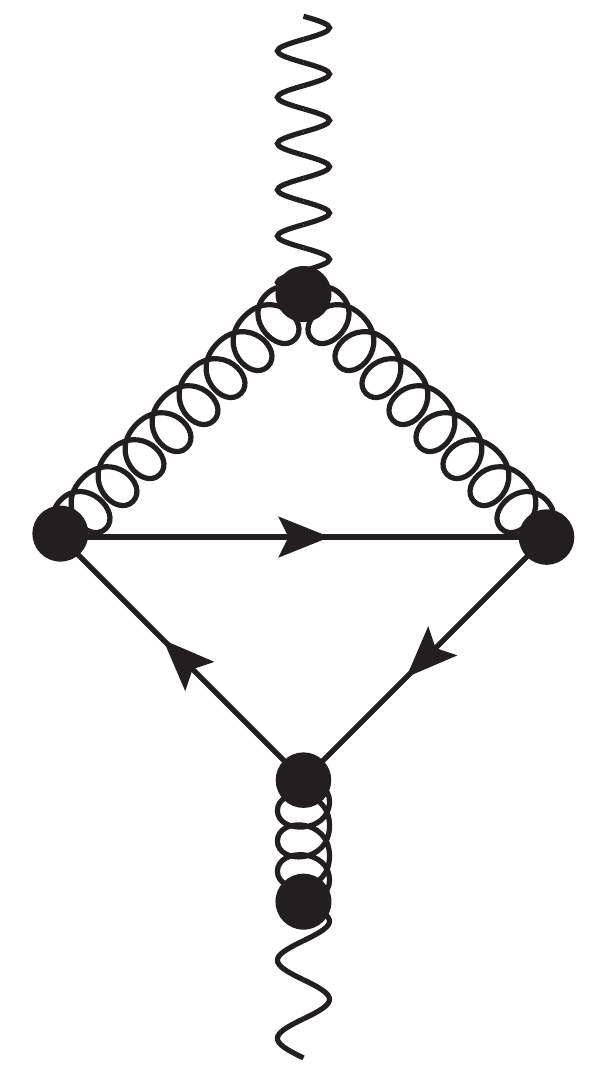}}
+
\parbox{1.4cm}{\vspace{0.1cm} \includegraphics[height = 2.5cm]{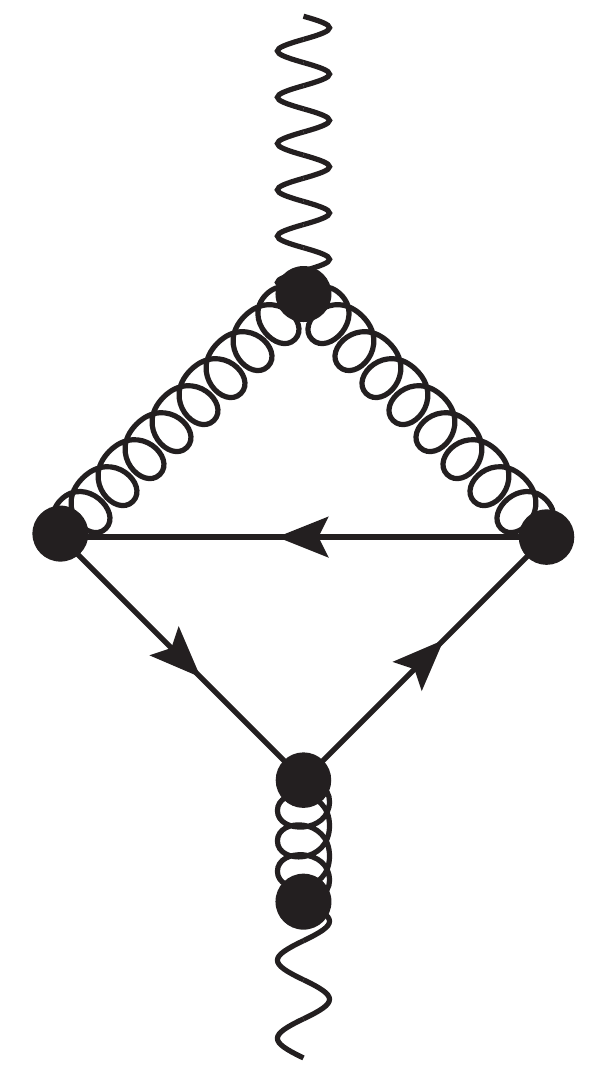}}
+
\parbox{1.4cm}{\vspace{0.1cm} \includegraphics[height = 2.5cm]{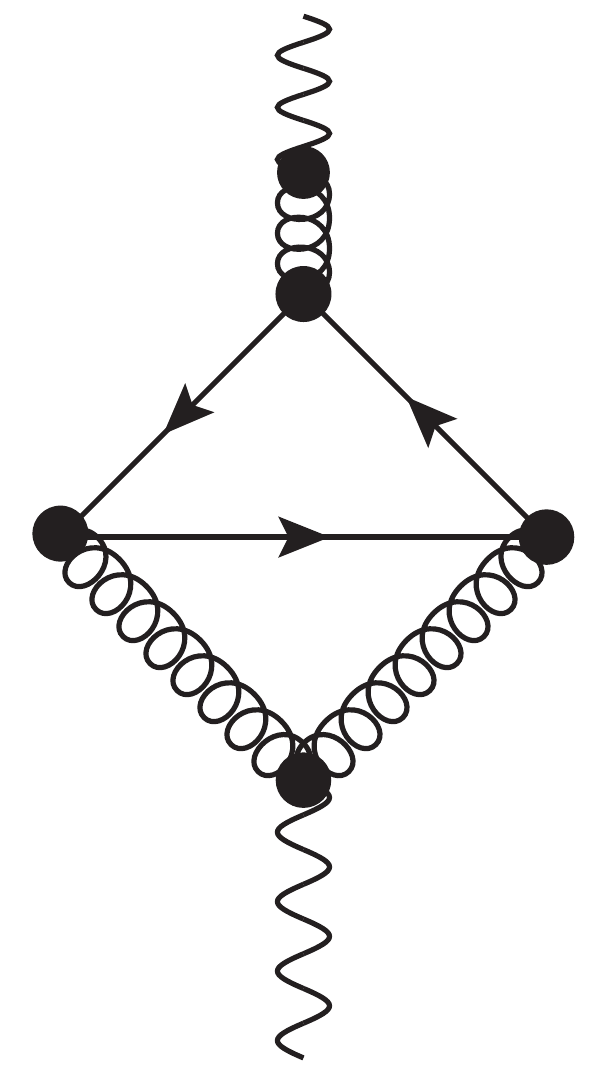}}
+
\parbox{1.4cm}{\vspace{0.1cm} \includegraphics[height = 2.5cm]{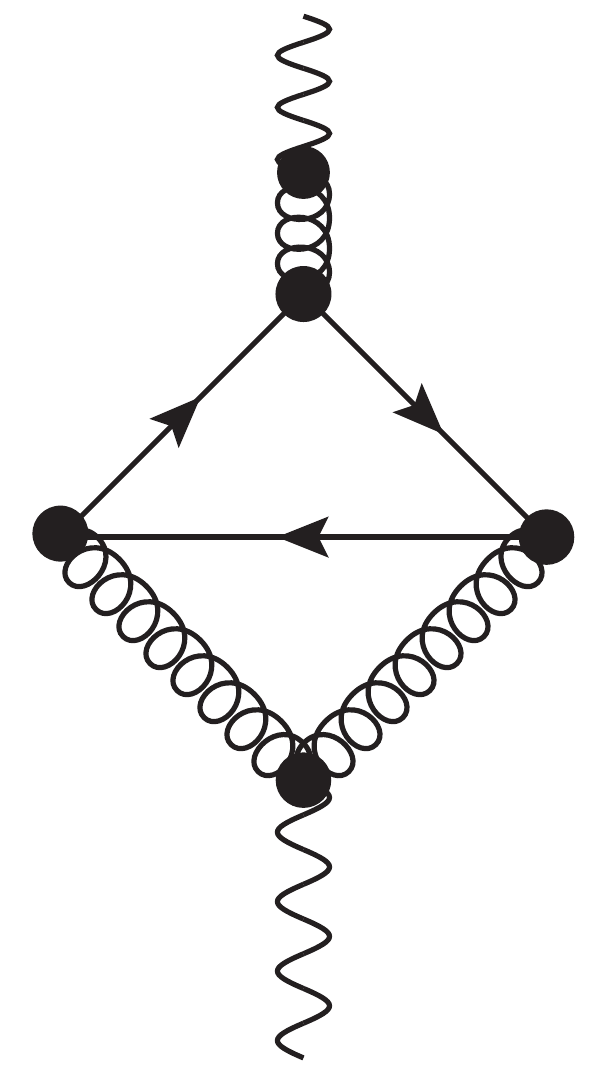}}
+
\parbox{1.4cm}{\vspace{0.1cm} \includegraphics[height = 2.5cm]{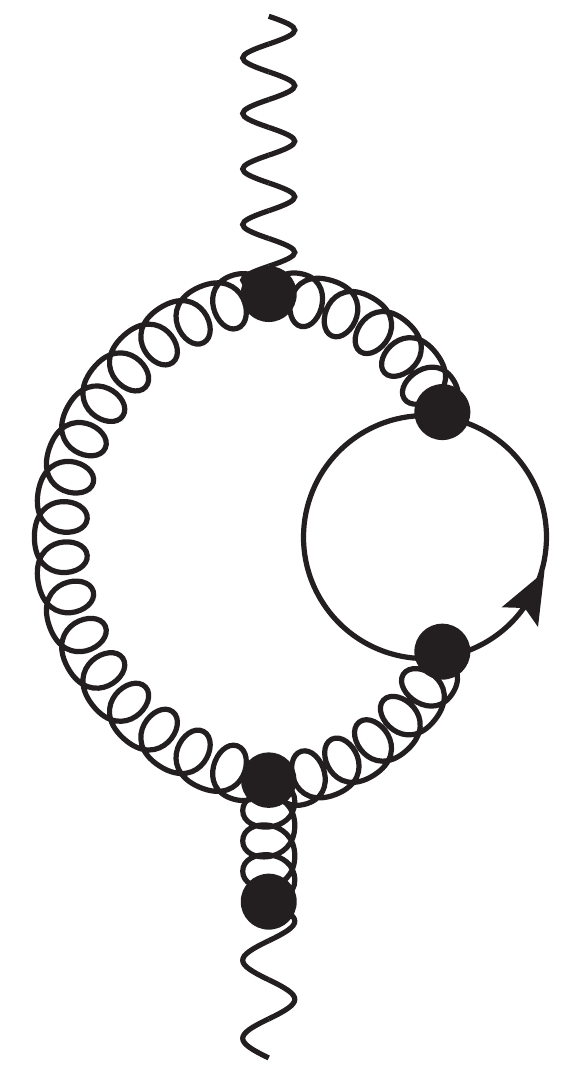}}
+
\parbox{1.4cm}{\vspace{0.1cm} \includegraphics[height = 2.5cm]{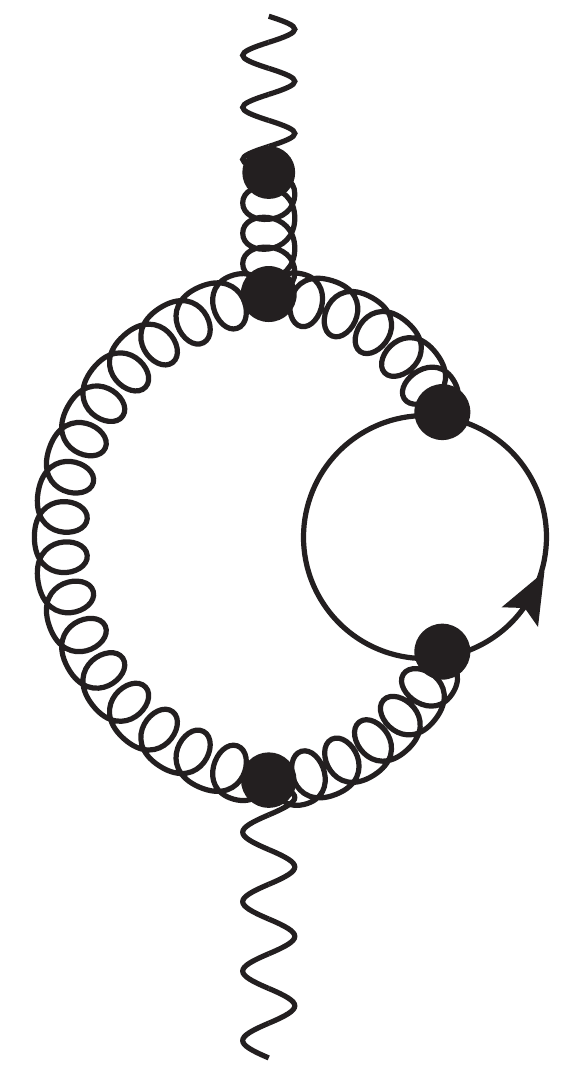}}
\parbox{1.5cm}{\vspace{0.1cm} $\,$}
+
\parbox{1.4cm}{\vspace{0.1cm} \includegraphics[height = 2.5cm]{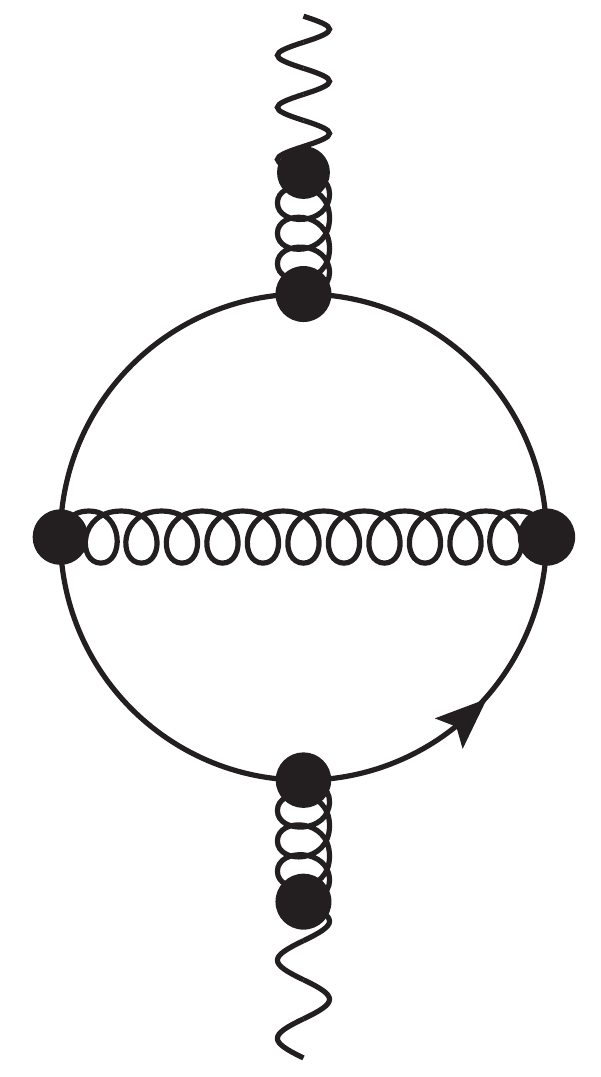}}
+
\parbox{1.4cm}{\vspace{0.1cm} \includegraphics[height = 2.5cm]{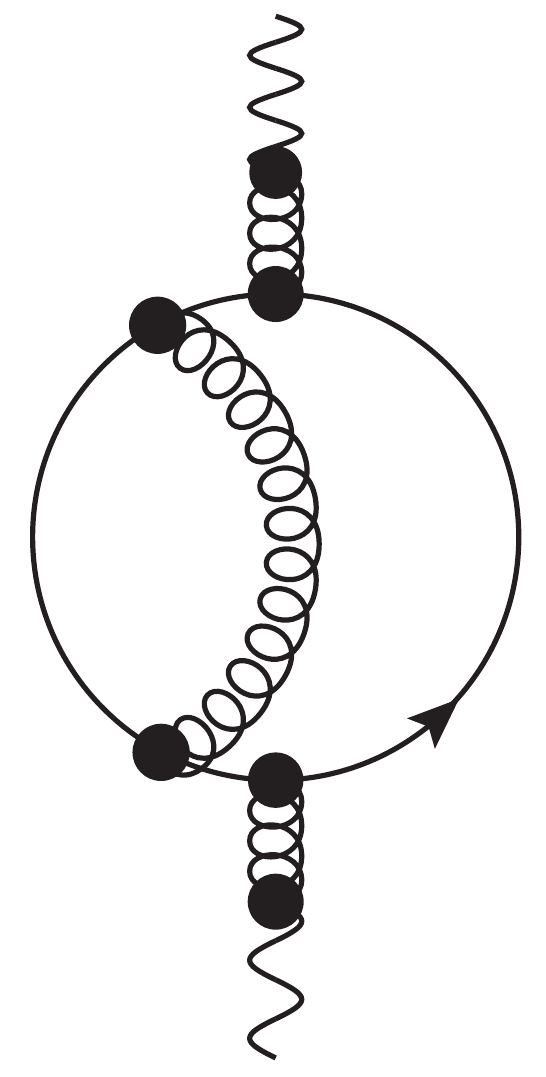}}
+
\parbox{1.4cm}{\vspace{0.1cm} \includegraphics[height = 2.5cm]{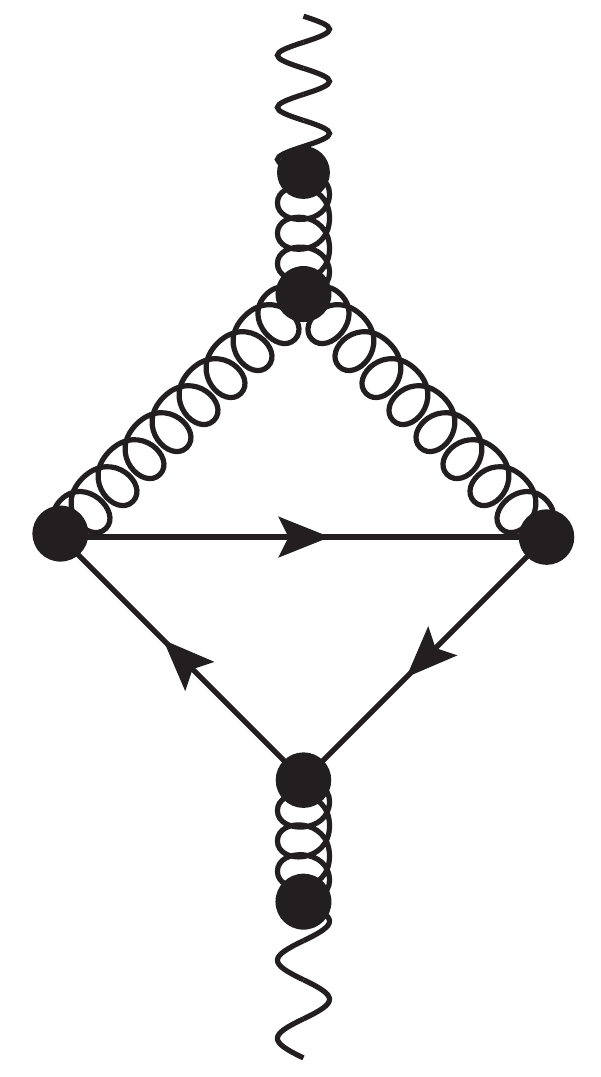}}
+
\parbox{1.4cm}{\vspace{0.1cm} \includegraphics[height = 2.5cm]{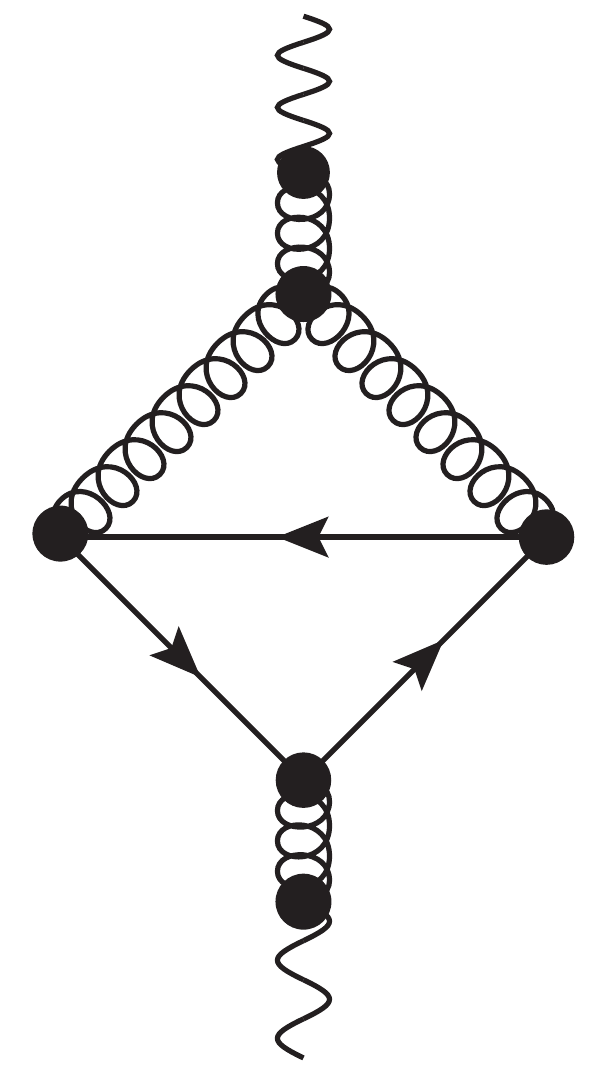}}
+
\parbox{1.4cm}{\vspace{0.1cm} \includegraphics[height = 2.5cm]{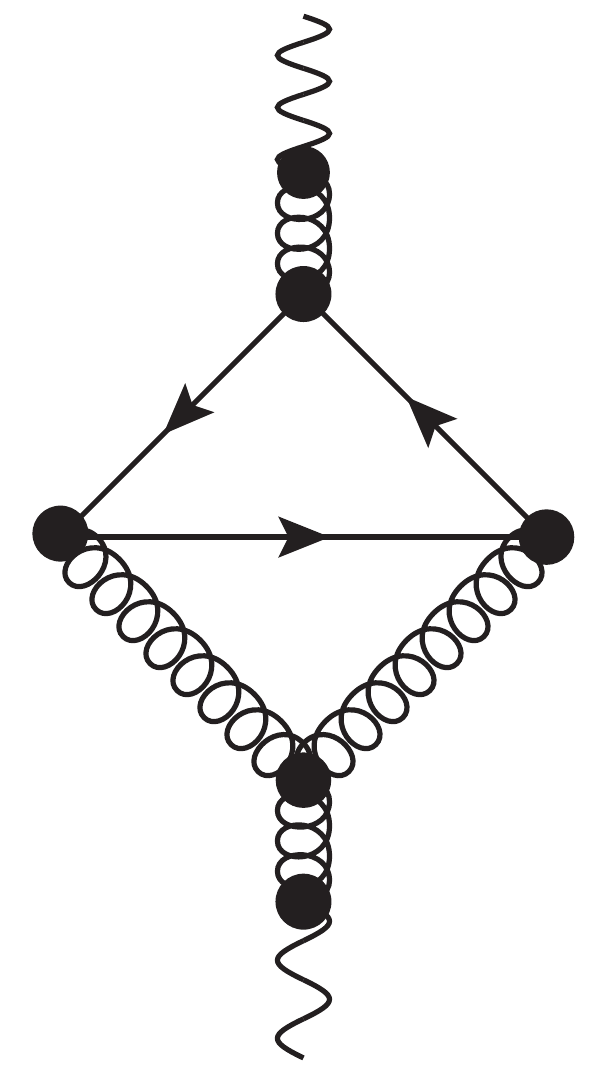}}
+
\parbox{1.4cm}{\vspace{0.1cm} \includegraphics[height = 2.5cm]{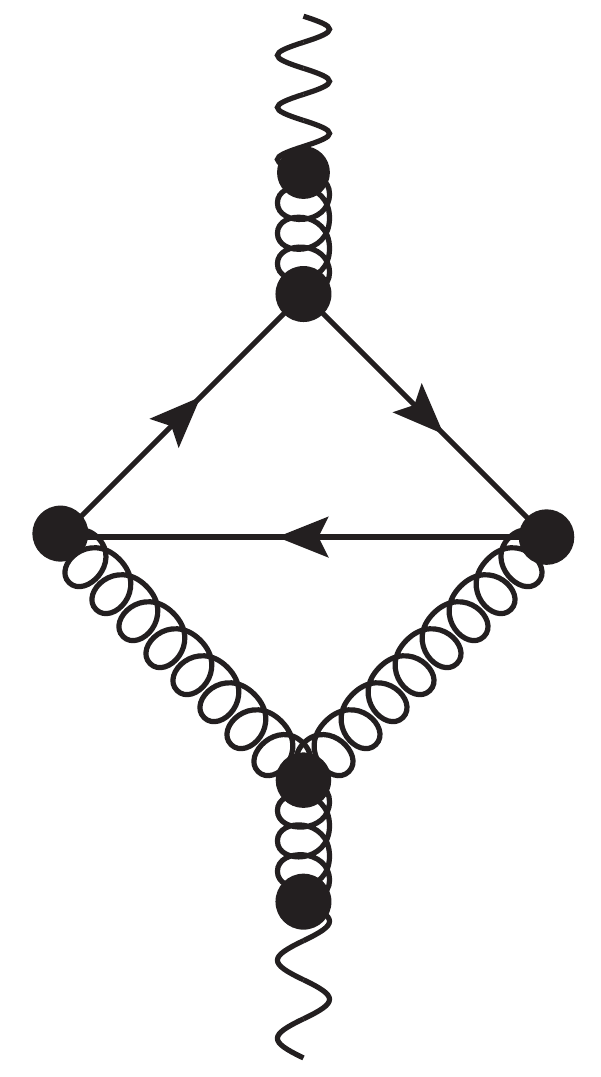}}
+
\parbox{1.4cm}{\vspace{0.1cm} \includegraphics[height = 2.5cm]{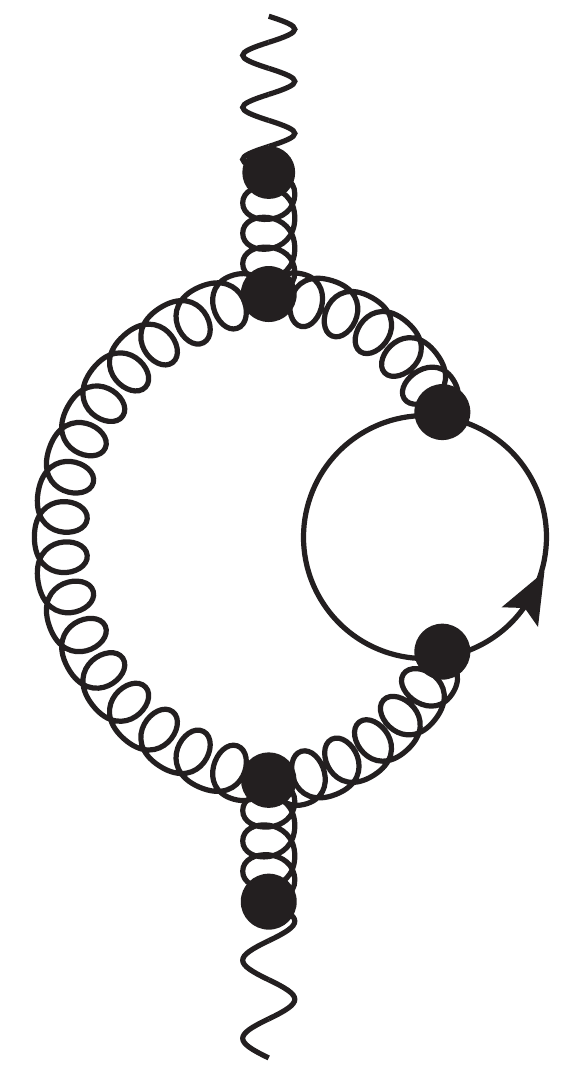}}
\caption{\small Diagrams contributing to the one-loop reggeized gluon self-energy.}
\label{fig:self_2loop}
\end{figure}
The complete set of contributing diagrams is given in
Fig.~\ref{fig:self_2loop}. The diagrams in the second line of this figure can be neglected since they 
do not give $\rho$-enhanced contributions. These subleading diagrams are obtained as projections of
the quark contribution to the two-loop gluon polarization tensor and are
therefore finite when $\rho \to \infty$.  

Among all the remaining diagrams we found that only the first one in Fig.~\ref{fig:self_2loop} is 
$\rho$-enhanced. This graph is unique since the reggeized gluon couples from above and below 
through an induced vertex to the usual gluon loop. For further technical details on this point we 
refer the reader to the Appendix. 

The complete set of enhanced contributions in Fig.~\ref{fig:self_2loop} is given by
\begin{align}
 \parbox{1.5cm}{\vspace{0.1cm} \includegraphics[height = 2.5cm]{qself2L.pdf}}   & =
-\rho  (-i 2 {\bm q}^2) \bar{g}^4 \frac{4 n_f}{\epsilon N_c}  \frac{\Gamma^2(2 + \epsilon)}{\Gamma(4 + 2\epsilon)} \cdot \frac{3 \Gamma(1 - 2\epsilon) \Gamma(1 + \epsilon) \Gamma(1 + 2\epsilon)}{\Gamma^2(1 - \epsilon) \Gamma(1 + 3 \epsilon) \epsilon}
  \label{eq:traj_jose}
\end{align}

It now necessary to introduce a subtraction procedure to avoid double counting of lower order contributions. For tree-level amplitudes, in quasi-multi-Regge-kinematics (QMRK),
such subtractions naturally arise as an alternative to impose explicit cut-offs on the rapidity of the produced particles, forcing them into strongly ordered clusters. In~\cite{Hentschinski:2011tz} it was shown that in the case of virtual corrections the subtracted pieces should contain both $\rho$-enhanced and non-$\rho$-enhanced terms.

In the work under discussion, the subtracted lower order contributions correspond to two one-loop self energies connected by a reggeized gluon propagator.  A diagramatic representation for the two-loop 
reggeized gluon self-energy would be as follows
\begin{align}
  \label{eq:coeff_2loop}
  \parbox{2cm}{\center \includegraphics[height = 2.5cm]{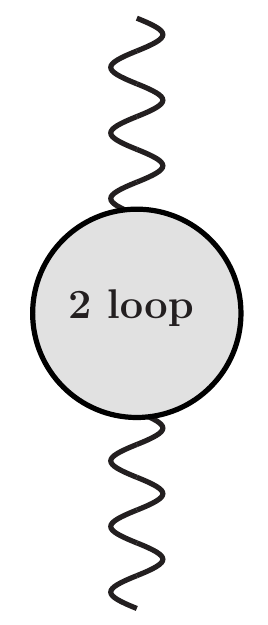}}  
  = 
  \parbox{2cm}{\center \includegraphics[height = 2.5cm]{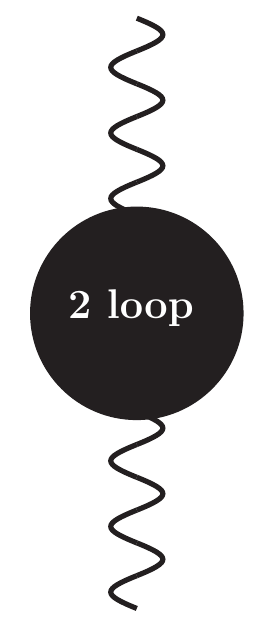}}
  -
  \parbox{2cm}{\center \includegraphics[height = 2.5cm]{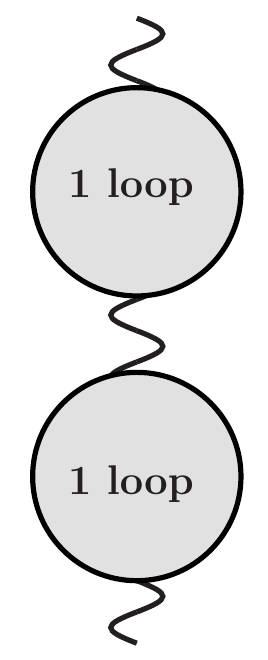}}.
\end{align}
More precisely, the terms to be subtracted which are both proportional to $n_f$ and $\rho$-enhanced read
\begin{align}
  \label{eq:doubleloo}
  \left[ \parbox{1.5cm}{\center \includegraphics[height = 2.5cm]{self2_1loop.pdf}}
\right]_{n_f, \,\rho}  
&= \hspace{0.4cm}
- \rho( -2i{\bm q}^2) \bar{g}^4  \left( \frac{{\bm q}^2}{\mu^2} \right)^{2 \epsilon} \frac{8 n_f}{\epsilon N_c} \frac{ \Gamma^2(2 + \epsilon) }{ \Gamma(4 + 2 \epsilon)}\frac{\Gamma^2(1 + \epsilon)}{ \Gamma(1 + 2\epsilon)}\frac{2}{\epsilon}.
\end{align}
The final subtracted reggeized gluon self energy, in terms of $n_f$ and $\rho$ contributions is
\begin{eqnarray}
 \Sigma^{(2)}_{n_f} \left(\rho; \epsilon, \frac{{\bm q}^2}{\mu^2}    \right) &=& 
 \left[   \parbox{1.5cm}{\center \includegraphics[height = 2.5cm]{self_2loop.pdf}} \right]_{n_f, \rho}  
 \nonumber\\
&&\hspace{-3.5cm}
   \begin{array}[h]{l}  
   \displaystyle  =  \frac{\rho( -2i{\bm q}^2) \bar{g}^4 4 n_f}{\epsilon N_c} \frac{ \Gamma^2(2 + \epsilon) }{ \Gamma(4 + 2 \epsilon)} \left( \frac{{\bm q}^2}{\mu^2} \right)^{2 \epsilon}  \left(\frac{\Gamma^2(1 + \epsilon)}{ \Gamma(1 + 2\epsilon)}\frac{4}{\epsilon}  -\frac{3 \Gamma(1 - 2\epsilon) \Gamma(1 + \epsilon) \Gamma(1 + 2\epsilon)}{\Gamma^2(1 - \epsilon) \Gamma(1 + 3 \epsilon) \epsilon}\right).
   \end{array}
   \label{eq:traje_quark_coeff}
\end{eqnarray}

\subsection{The quark contribution to the two-loop reggeized gluon trajectory}
\label{sec:trajecto}

In order to obtain from the previous results the quark contribution to the two-loop reggeized 
gluon trajectory, it is needed to explicitly construct the high energy limit of a given partonic QCD 
scattering amplitude at two loops in the next-to-leading logarithmic accuracy.  Besides the two-loop correction to the reggeized gluon self-energy, just discussed, we also need the one-loop partonic 
impact factors. Within the effective action approach only the quark impact factor is already known at
one-loop accuracy. It is then convenient to consider elastic quark-quark scattering $p_a + p_b \to p_1 + p_2$ as our test amplitude to extract the trajectory.  

Using Lipatov's effective action, at one loop the high energy limit of the quark-quark scattering 
amplitude is given by the sum~\cite{Hentschinski:2011tz}:
\begin{align}   
i\mathcal{M}_{q_a q_b \to q_1 q_2}^{(1)} =
\parbox{2cm}{\includegraphics[width = 2cm]{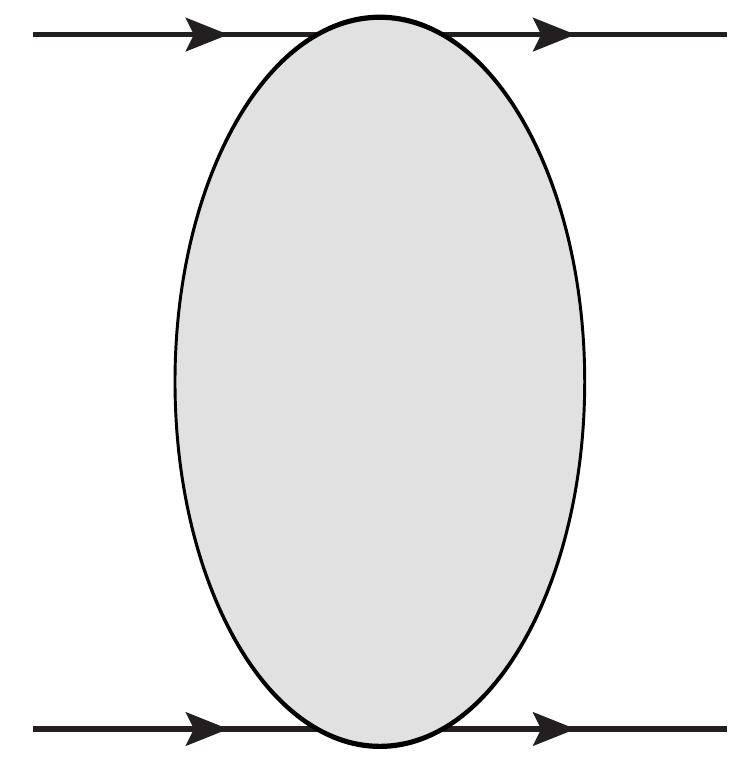}}
&= 
\parbox{2cm}{\includegraphics[width = 2cm]{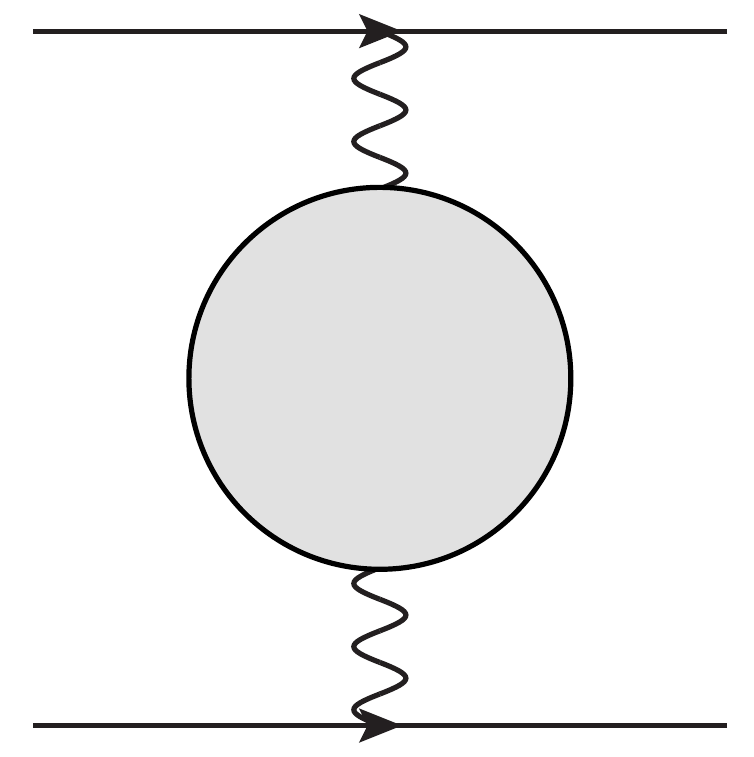}}
+
\parbox{2cm}{\includegraphics[width = 2cm]{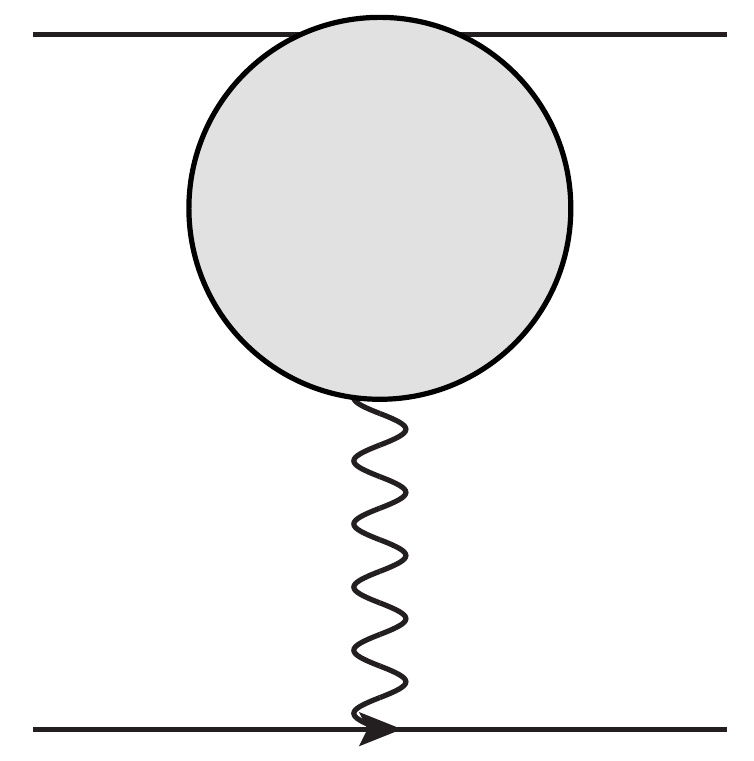}}
+
\parbox{2cm}{\includegraphics[width = 2cm]{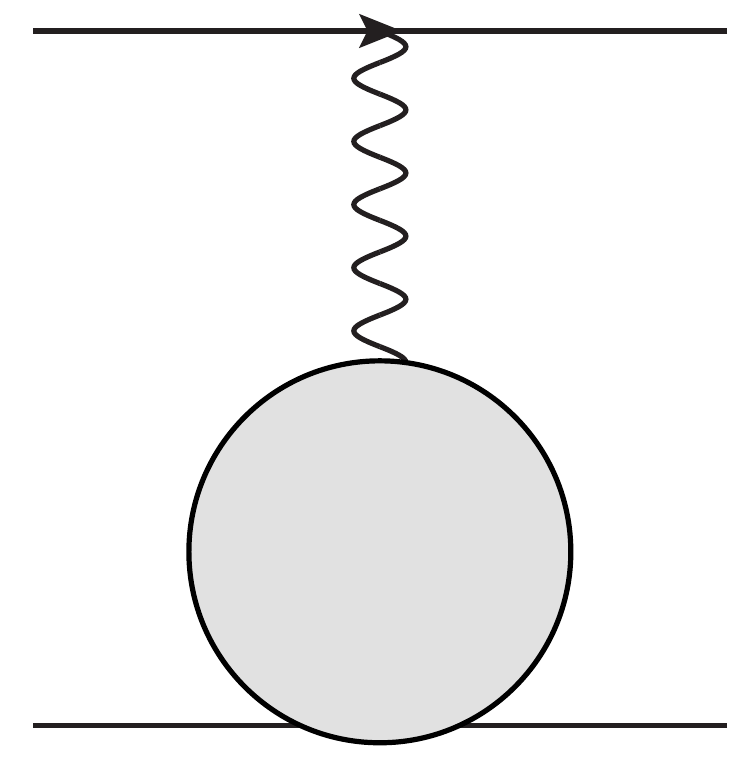}}. 
\label{eq:1loopgra}
\end{align}
While the one-loop reggeized self-energy is given by Eq.~\eqref{eq:self_1loop}, the subtracted one-loop quark-reggeized gluon coupling reads~\cite{Hentschinski:2011tz}
\begin{align}\label{eq:coeff_impa}
 \mathcal{C}^{(1)}_{qr^* \to q} & \left(  p_a^+, \rho; \epsilon \frac{{\bm q}^2}{\mu^2} \right) = \!\!\parbox{1.6cm}{\includegraphics[width = 2cm]{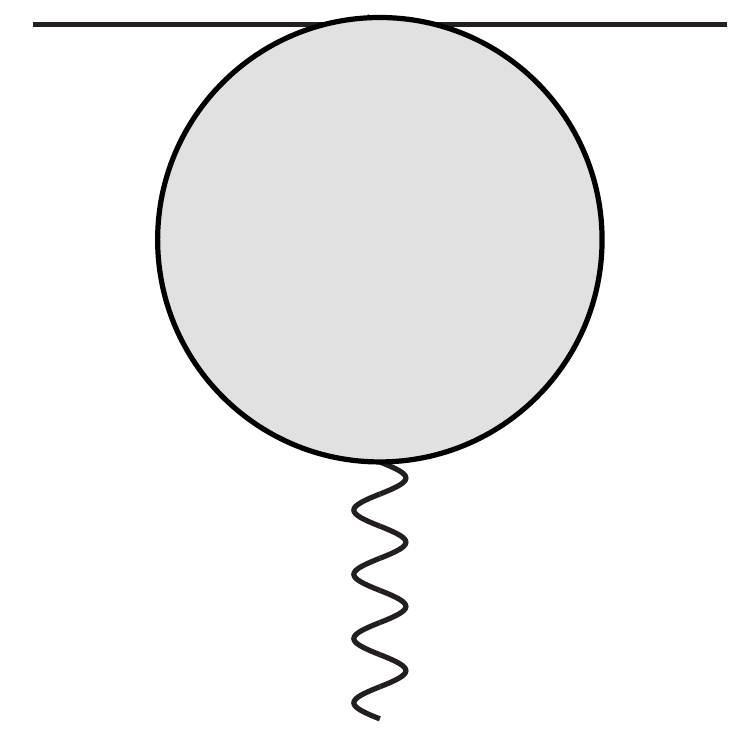}} =
  i \mathcal{M}^{(0)}_{qr^* \to q}  \bar{g}^2 
% \frac{g^2  }{(4\pi)^{2 + \epsilon}}  
\left(\frac{ {\bm q}^2}{\mu^2} \right)^\epsilon  \frac{ \Gamma^2 (1+\epsilon)}{\Gamma(1+2 \epsilon)}  \left\{
 \frac{-2 }{\epsilon}  
\left(  \ln \frac{p_a^+}{\sqrt{{\bm q}^2}} - \frac{\rho}{2} \right)  \right. \notag \\
& \left. +
\frac{  (2 + 7 \epsilon)}{2\epsilon^2(1 + 2\epsilon)} + 
 \frac{1}{N^2_c} 
\left(  \frac{1}{\epsilon^2 (1 + 2 \epsilon)} + \frac{1}{2 \epsilon} \right) 
 - 
\frac{1}{\epsilon}  \bigg(\psi(1-\epsilon) - 2 \psi(\epsilon) + \psi(1)\bigg)\right\},
\end{align}
with $ i \mathcal{M}^{(0)}_{qr^* \to q} = igt^a p_a^+ $ being the quark
-reggeized gluon coupling at Born level. 

Each of the three diagrams at the right hand side of Eq.~\eqref{eq:1loopgra} is divergent in the 
$\rho \to \infty$ limit. However, these $\rho$-divergencies cancel in the sum, yielding the following result:
\begin{align}\label{eq:1loopres}
i\mathcal{M}_{q_a q_b \to q_1 q_2}^{(1)} &=  i\mathcal{M}_{q_a q_b \to q_1 q_2}^{(0)}  
\bigg[\frac{1}{2} \left( \ln \frac{s}{ {\bm q}^2}  +  \ln \frac{-s}{ {\bm q}^2 } \right) \omega^{(1)}(- {\bm q}^2)  
+ \Gamma_a^{(1)} ( {\bm q}^2) + \Gamma_b^{(1)} ( {\bm q}^2) \bigg].
\end{align}
Here $ \omega^{(1)}(- {\bm q}^2) $ denotes the one-loop gluon Regge 
trajectory, while $\Gamma_{a,b}^{(1)} ( {\bm q}^2)$ are the one-loop
quark impact factors which arise from the combination of finite terms in Eq.~\eqref{eq:self_1loop}
and Eq.~\eqref{eq:coeff_impa}. 

The cancellation of $\rho$-divergencies can be interpreted as a consequence of high energy factorization. While the scattering amplitude is free of high energy divergencies, both the reggeized gluon-quark 
coupling and the reggeized gluon propagator carry divergencies from the one-loop level on. 
This fact resembles the similar situation of collinear factorization where both
partonic cross sections and parton distribution functions carry divergencies starting from  
the one-loop level which cancel when are put together. 

It is interesting to approach the calculation of the gluon trajectory from the point of view of renormalized 
effective vertices and propagators. Formally, this can be understood as a renormalization of the 
coefficients of the reggeized gluon action as obtained after 
integrating out gluon and quark fields with the subsequent subtraction of non-local contributions. At the amplitude level this corresponds to the following definition of the renormalized quark-reggeized gluon coupling coefficients:
\begin{align}
  \label{eq:renomcoeff}
   \mathcal{C}^{ \text{R}}_{qr^* \to q} \left( \frac{ p_a^+}{M^+}; \epsilon,  \frac{{\bm q}^2}{\mu^2} \right) 
& =
 Z^+ \left( \frac{M^+}{\sqrt{{\bm q}^2}}, \rho ; \epsilon,  \frac{{\bm q}^2}{\mu^2}  \right)   \mathcal{C}_{qr^* \to q} \left(  \frac{ p_a^+}{\sqrt{{\bm q}^2}}, \rho; \epsilon \frac{{\bm q}^2}{\mu^2} \right), 
\\
   \mathcal{C}^{ \text{R}}_{qr^* \to q} \left( \frac{ p_b^-}{M^-}; \epsilon,  \frac{{\bm q}^2}{\mu^2} \right) 
& =
 Z^- \left( \frac{M^-}{\sqrt{{\bm q}^2}}, \rho ; \epsilon,  \frac{{\bm q}^2}{\mu^2}  \right)   \mathcal{C}_{qr^* \to q} \left(  \frac{ p_b^-}{\sqrt{{\bm q}^2}}, \rho; \epsilon \frac{{\bm q}^2}{\mu^2} \right), 
\end{align}
and the renormalized reggeized gluon propagator:
\begin{align}
  \label{eq:renompropR}
  G^{\text{R}} \left(M^+, M^-; \epsilon, {\bm q}^2, \mu^2 \right)  & = \frac{ G \left(\rho; \epsilon, {\bm q}^2, \mu^2   \right) }{ Z^+ \left( \frac{M^+}{\sqrt{{\bm q}^2}}, \rho ; \epsilon,  \frac{{\bm q}^2}{\mu^2}  \right) Z^- \left( \frac{M^-}{\sqrt{{\bm q}^2}}, \rho ; \epsilon,  \frac{{\bm q}^2}{\mu^2}  \right) },
\end{align}
with the bare reggeized gluon propagator given by
\begin{align}
  \label{eq:barepropR}
   G \left(\rho; \epsilon, {\bm q}^2, \mu^2   \right)
&=
\frac{i/2}{{\bm q}^2} \left\{ 1 + \frac{i/2}{{\bm q}^2} \Sigma \left(\rho; \epsilon, \frac{{\bm q}^2}{\mu^2}    \right)  + \left[  \frac{i/2}{{\bm q}^2} \Sigma \left(\rho; \epsilon, \frac{{\bm q}^2}{\mu^2}   \right)\right] ^2 + \ldots   \right\}. 
\end{align}
The renormalization factors $Z^\pm$ cancel for the complete quark-quark scattering amplitude and can be  parameterized as follows
\begin{align}
  \label{eq:Z+-}
  Z^\pm\left( \frac{M^\pm}{\sqrt{{\bm q}^2}}, \rho ; \epsilon,  \frac{{\bm q}^2}{\mu^2}  \right)  & = \exp \left [ \left(\rho -  \ln \frac{M^\pm}{\sqrt{\bm q}^2}  \right) \omega\left(\epsilon, \frac{{\bm q}^2}{\mu^2} \right) + f\left(\epsilon, \frac{{\bm q}^2}{\mu^2} \right) \right]
\end{align}
where the gluon Regge trajectory has the following perturbative expansion:
\begin{align}
  \label{eq:omega_expand}
   \omega\left(\epsilon, \frac{{\bm q}^2}{\mu^2} \right) & =  \omega^{(1)}\left(\epsilon, \frac{{\bm q}^2}{\mu^2} \right) + \omega^{(2)}\left(\epsilon, \frac{{\bm q}^2}{\mu^2} \right) + \ldots
\end{align}
At one loop $ \omega({\bm q}^2) $ coincides (up to the overall factor $-2
i {\bm q}^2$, which cancels against the adjacent reggeized gluon
propagators in the amplitude) with the term proportional to $\rho$ in the one-loop
reggeized gluon self-energy found in Eq.~\eqref{eq:self_1loop}, {\it i.e.}
\begin{align}
  \label{eq:omega1}
  \omega^{(1)}\left(\epsilon, \frac{{\bm q}^2}{\mu^2} \right) & = -\frac{2 \bar{g}^2 \Gamma^2(1 + \epsilon)}{\Gamma(1 + 2 \epsilon)\epsilon } \left(\frac{{\bm q}^2}{\mu^2} \right)^\epsilon.  
\end{align}

The function $f(\epsilon, {\bm k}^2)$ parameterizes finite contributions and is, 
in principle, arbitrary.  Regge theory suggests to fix it in such a way that at one loop the non-$\rho$-enhanced contributions of the one-loop reggeized gluon self energy are entirely transferred to the
quark-reggeized gluon couplings leading to 
\begin{align}
  \label{eq:f1loop}
   f^{(1)}\left(\epsilon, \frac{{\bm q}^2}{\mu^2} \right) & =  \frac{ \bar{g}^2 \Gamma^2(1 + \epsilon)}{\Gamma(1 + 2 \epsilon)} \left(\frac{{\bm q}^2}{\mu^2} \right)^\epsilon  
         \frac{(-1)}{(1 + 2 \epsilon)2 \epsilon} \bigg[   2 
 +  \frac{5 + 3\epsilon}{3 + 2 \epsilon} 
-\frac{n_f}{N_c} \left( \frac{2 + 2\epsilon}{3 + 2\epsilon} \right)\bigg]  .
\end{align}
Using $M^+ = p_a^+$, $M^- = p_b^-$ we can see that this choice of $f$ keeps the full $s$-dependence of 
the amplitude  inside the reggeized gluon exchange. Furthermore, the renormalized
quark-reggeized gluon couplings then agree with the one-loop quark impact
factors originally calculated in~\cite{Fadin:1993qb}, and derived again in~\cite{DelDuca:1998kx} 
and~\cite{Hentschinski:2011tz}. 

To extract the quark pieces of the gluon Regge trajectory at two loops it is needed to identify 
the $\rho$-enhanced contributions of the bare reggeized gluon propagator up to two loops. These are 
\begin{align}
  \label{eq:GBare2loop}
   G \left(\rho; \epsilon, {\bm q}^2, \mu^2   \right) & = \frac{i/2}{{\bm q}^2} \bigg\{ 1 + \frac{i/2}{{\bm q}^2} \Sigma^{(1)} \left(\rho; \epsilon, \frac{{\bm q}^2}{\mu^2}    \right) 
\notag \\
&  
+ 
\frac{i/2}{{\bm q}^2} \Sigma^{(2)} \left(\rho; \epsilon, \frac{{\bm q}^2}{\mu^2}    \right)
+
\left[  \frac{i/2}{{\bm q}^2} \Sigma^{(1)} \left(\rho; \epsilon, \frac{{\bm q}^2}{\mu^2}  \right)   \right]^2 + \ldots   \bigg\}. 
\end{align}
Selecting only the terms proportional to $n_f$, the pieces of ${\cal O} (\rho)$ in the renormalized 
reggeized gluon propagator at two loops are
\begin{align}
  \label{eq:GRrho}
   G^{\text{R}} \bigg|_{n_f, \rho}
 & =  \frac{i/2}{{\bm q}^2} \bigg\{  \frac{i/2}{{\bm q}^2} \Sigma^{(2)}_{n_f} +  
\left[ 
 \frac{i/2}{{\bm q}^2} 
\Sigma^{(1)} \left(\rho;\epsilon, \frac{{\bm q}^2}{\mu^2} \right)   \right]^2 \bigg|_{n_f, \rho}
\notag \\
& 
- \rho \,  \omega^{(1)}\left(\epsilon, \frac{{\bm q}^2}{\mu^2} \right) f^{(1)}_{n_f} \left(\epsilon, \frac{{\bm q}^2}{\mu^2} \right) -  \rho \, \omega^{(2)}_{n_f}\left(\epsilon, \frac{{\bm q}^2}{\mu^2} \right) \bigg\}
\end{align}
with
\begin{align}
  \label{eq:fnf}
  f^{(1)}_{n_f} \left(\epsilon, \frac{{\bm q}^2}{\mu^2} \right) & = \frac{ \bar{g}^2 \Gamma^2(1 + \epsilon)}{\Gamma(1 + 2 \epsilon)} \left(\frac{{\bm q}^2}{\mu^2} \right)^\epsilon  \frac{n_f}{N_c}  
  \bigg(\frac{2 + 2\epsilon}{3 + 2\epsilon}\bigg) 
\end{align}
being the part proportional to $n_f$ of $f^{(1)}$ in Eq.~\eqref{eq:f1loop}. 

The requirement that the $\rho$ dependence in Eq.~\eqref{eq:GRrho} has to cancel in the limit 
$\rho \to \infty$ then yields the quark contribution to the gluon Regge trajectory at two loops as
\begin{align}
  \label{eq:traje_quark_coeff}
 \omega^{({2})}_{n_f}\left(\epsilon, \frac{{\bm q}^2}{\mu^2} \right)  & =  \bar{g}^4  \left( \frac{{\bm q}^2}{\mu^2} \right)^{2 \epsilon} \frac{4 n_f}{\epsilon N_c} \frac{ \Gamma^2(2 + \epsilon) }{ \Gamma(4 + 2 \epsilon)} \bigg [ \frac{\Gamma^2(1 + \epsilon)}{ \Gamma(1 + 2\epsilon)}\frac{2}{\epsilon} -\frac{3 \Gamma(1 - 2\epsilon) \Gamma(1 + \epsilon) \Gamma(1 + 2\epsilon)}{\Gamma^2(1 - \epsilon) \Gamma(1 + 3 \epsilon) \epsilon}\bigg],
\end{align}
which is in perfect agreement with Eq.(9) of the original study of Fadin et al. in~\cite{Fadin:1996tb}. 
This confirms the validity of our approach 
to the effective action at two loop level and the regularization prescription here presented. The 
application of this approach to calculate the gluon contributions in the two-loop gluon Regge 
trajectory is the subject of our current investigations.

\section{Conclusions}

The first computation of a two-loop result using Lipatov's high energy effective action has been presented, 
finding perfect agreement with the results previously derived by Fadin et al~\cite{Fadin:1993qb} for the quark 
contribution to the two-loop gluon Regge trajectory. The regularization and subtraction prescriptions introduced 
in~\cite{Hentschinski:2011tz} have been shown to be valid also at two loops. Work is in progress to apply similar 
methods in the more complicated case of the gluon contributions both for the forward jet vertex calculation, as 
in~\cite{Hentschinski:2011tz}, and for the two-loop gluon Regge trajectory, as in the present work. The derivation of the NLO quark Regge trajectory is also ongoing work. The successful 
completion of this program will open the possibility to make use of this powerful formalism for collider phenomenology. 
The study of the implications of the high energy effective action for scattering amplitudes in 
supersymmetric theories~\cite{SUSY}, gravity~\cite{gravity} and integrable systems~\cite{Romagnoni:2011jr} is also underway.

\subsubsection*{Acknowledgements}\label{5}
Discussions with J.~Bartels and L.N.~Lipatov are acknowleged.  G. C. thanks the Department of Theoretical Physics at the Aut{\' o}noma University of Madrid and the ÒInstituto de F{\' \i}sica Teorica UAM/CSICÓ for their hospitality. Research partially funded by the German Academic Exchange Service (DAAD),
MICINN under grant FPA-2009-07908, European Comission (LHCPhenoNet
PITN-GA-2010-264564) and Comunidad de Madrid (HEPHACOS ESP-1473).
\appendix

\section{Evaluation of Feynman diagrams}
\label{sec:details}

In the following some details related to the calculation of the Feynman 
diagrams in the first line of Fig.~\ref{fig:self_2loop} are given. Longitudinal
divergencies are regularized by deforming the light-cone vectors of the
effective action according to Eq.~\eqref{eq:n+-} and all integrals are evaluated in the limit $\rho \to \infty$.  
Ultraviolet, soft and collinear divergencies are regularized using dimensional
regularization with $d = 4 + 2 \epsilon$. For some of the diagrams
integration-by-parts (IBP) relations have been used and the integrand has been
reduced to master integrals making use of the implementation of the
Laporta algorithm~\cite{Laporta:2001dd} in FIRE~\cite{Smirnov:2008iw}.
\\
\\
Following these techniques, the result for the only $\rho$-enhanced diagram can be written as
\begin{align}
  \label{eq:enhanced}
  \parbox{1.4cm}{\vspace{0.1cm} \includegraphics[height = 2.5cm]{qself2L_d1.pdf}}
  \begin{array}[h]{l}
    \\ \\ \displaystyle 
 = \hspace{.4cm} 
 \frac{i 4 {\bm q}^2 g^4 N_c}{(4 \pi)^{4 + 2\epsilon}} \left(\frac{{\bm q}^2}{\mu^2} \right)^{2\epsilon} 
\frac{\Gamma(\epsilon) \Gamma(1-\epsilon) \Gamma(2 + \epsilon)}{(3 + 2\epsilon) \Gamma(2 + 2\epsilon)} 
\\ \\ 
\qquad \qquad \qquad \times  \displaystyle 
\left( 2 \rho  \frac{\Gamma(1-2\epsilon) \Gamma(\epsilon) \Gamma(2\epsilon)}{\Gamma(1-\epsilon)\Gamma(3\epsilon)}  -  \frac{\Gamma(1 + \epsilon) \Gamma(2\epsilon) \Gamma(1 - 2 \epsilon)}{\Gamma(2 - \epsilon) \Gamma(1 + 3\epsilon)} \right).
  \end{array}
\end{align}
Note that the term proportional to $\rho$ is a different representation of the same result in Eq.~(\ref{eq:traj_jose}). Two other types of diagrams in Fig.~\ref{fig:self_2loop} are potentially $\rho$-enhanced. However, a detailed calculation shows that this is not the case. The first of these diagrams is the combination
\begin{align}
  \label{eq:Sc1}
{S}_{\rm (c_1)} & =   \parbox{1.4cm}{\vspace{0.1cm} \includegraphics[height = 2.5cm]{qself2L_c1i.pdf}}
+
\parbox{1.4cm}{\vspace{0.1cm} \includegraphics[height = 2.5cm]{qself2L_c2i.pdf}}
 = -\frac{g^4N_c}{2S}(\mu^2)^{-2\epsilon}{S}_1,
\end{align}
where
\begin{equation}
\begin{aligned}
&\hspace{-0.3cm}{S}_1=\iint\frac{d^dl}{(2\pi)^d}\frac{d^dk}{(2\pi)^d}\frac{{\rm Tr}[\slashed{n_a}(\slashed{q}+\slashed{l})\slashed{n_b}\slashed{l}  \slashed{n}_a 
(\slashed{l}-\slashed{k})]}{k\cdot n_a\, k^2 l^2(k+q)^2(l+q)^2(k-l)^2}.
\end{aligned}
\end{equation}
%with the symmetry factor $S=2$.  
%The evaluation of the integral ${S}_1$ requires a reduction to master integrals by means of IBP relations for which we used FIRE~\cite{Smirnov:2008iw}.  
The result for ${S}_1$ can then be given in terms of three master integrals:
\begin{eqnarray}
S_1 &=& 4n_a \cdot n_b\left[\frac{2}{\mathbf{q}^2}\left(\frac{(3+4\epsilon)(2\epsilon^2+7\epsilon+2)}{(1+2\epsilon)(3+2\epsilon)}K_1-\frac{12\epsilon^5+36\epsilon^4+27\epsilon^3-20\epsilon^2-26\epsilon-6}{\epsilon^2(1+2\epsilon)(3+2\epsilon)^2}K_2\right) \nonumber \right.\\
&&\left. +\frac{1}{3+2\epsilon}K_3\right].
\end{eqnarray}
%\begin{equation}
%\begin{aligned}
%{S}_1&=8\frac{a\cdot b}{\vec{q}^2}\frac{3+4\epsilon}{\epsilon}\left[\frac{1+4\epsilon}{2(1+2\epsilon)}+\frac%{1+2\epsilon}{3+2\epsilon}\right]
%{K}_1\\
%&+2\frac{n_a\cdot n_b}{{\bm q}^2}\frac{1}{\epsilon^2}\bigg[\frac{(1+3\epsilon)(2+3\epsilon)(1-2\epsilon)}%{(1+2\epsilon)}+\frac{4(1+2\epsilon)}{(3+2\epsilon)^2}(2\epsilon^2+11\epsilon+6)\bigg]{K}_2\\
%&+2n_a\cdot n_b\left[1+\frac{4}{\epsilon}-\frac{6(1+2\epsilon)(3+\epsilon)}{\epsilon (3+2\epsilon)^2}\right]{K}_3,
%\end{aligned}
%\end{equation}
with
\begin{equation}\label{mast}
\begin{aligned}
{K}_1&=\iint\frac{d^dl}{(2\pi)^d}\frac{d^dk}{(2\pi)^d}\frac{n_a\cdot l}{n_a\cdot k \,l^2(l+q)^2(k-l)^2} \notag \\
& \qquad \qquad \qquad \qquad  = 
\frac{1}{(4\pi)^{4+2\epsilon}} \frac{\Gamma^2(-1-2\epsilon)\Gamma^2(2+2\epsilon)\Gamma(1+\epsilon)\Gamma(-1-\epsilon)}{\Gamma(4+4\epsilon)\Gamma(-2-2\epsilon)}({\bm q}^2)^{1+2\epsilon},
\\
{K}_2&=\iint\frac{d^dl}{(2\pi)^d}\frac{d^dk}{(2\pi)^d}\frac{1}{k^2(l+q)^2(k-l)^2}
 = \frac{({\bm q}^2)^{1+2\epsilon}}{(4\pi)^{4+2\epsilon}}\frac{\Gamma^3(1+\epsilon)\Gamma (-1-2\epsilon)}{\Gamma (3+3\epsilon)},
\\
{K}_3&=\iint\frac{d^dl}{(2\pi)^d}\frac{d^dk}{(2\pi)^d}\frac{1}{k^2l^2(k+q)^2(l+q)^2} = -\frac{({\bm q}^2)^{2\epsilon}}{(4\pi)^{4+2\epsilon}}\left[\frac{\Gamma^2 (1+\epsilon)\Gamma (-\epsilon)}{\Gamma (2+2\epsilon)}\right]^2,
\end{aligned}
\end{equation}
which are all finite in the limit $\rho \to \infty$. The same is true for diagrams of the type 
\begin{align}
  \label{eq:d2}
 {S}_{\rm (d_2)} & =   \parbox{1.4cm}{\vspace{0.1cm} \includegraphics[height = 2.5cm]{qself2L_d2iu.pdf}} = -\frac{2i^{1-2\epsilon}}{(4\pi)^{2+\epsilon} S}{\bm q}^2 N_c \frac{\Gamma (\epsilon)\Gamma (1-\epsilon) \Gamma (2+\epsilon)}{(3+2\epsilon)\Gamma (2+2\epsilon)}\left({C}_1-n_a\cdot n_b \,{C}_2 \right),
\end{align}
where
\begin{equation}\label{c}
\begin{aligned}
{C}_1&=\int\frac{d^dk}{(2\pi)^d}\frac{k\cdot n_a\, k\cdot n_b}{k^2[(k-q)^2]^{2-\epsilon}}=\frac{i^{1+2\epsilon}}{2(4\pi)^{2+\epsilon}}({\bm q}^2)^{2\epsilon}n_a\cdot n_b\frac{\Gamma (1+2\epsilon)\Gamma (2+\epsilon)\Gamma (-2\epsilon)}{\Gamma (2-\epsilon)\Gamma (3+3\epsilon)},\\
{C}_2&=\int\frac{d^dk}{(2\pi)^d}\frac{k\cdot (k-q)}{k^2[(k-q)^2]^{2-\epsilon}}=\frac{i^{-1+2\epsilon}}{(4\pi)^{2+\epsilon}}({\bm q}^2)^{2\epsilon}\frac{\Gamma (2\epsilon)\Gamma (2+\epsilon)\Gamma (1-2\epsilon)}{\Gamma (2-\epsilon)\Gamma (2+3\epsilon)}.
\end{aligned}
\end{equation}


\begin{thebibliography}{99}

%\cite{Hentschinski:2011tz}
\bibitem{Hentschinski:2011tz}
  M.~Hentschinski, A.~Sabio~Vera,
% ``NLO jet vertex from Lipatov's QCD effective action'', 
  arXiv:1110.6741 [hep-ph].
  %%CITATION = ARXIV:1110.6741;%%  
  
 \bibitem{BFKL1}  
L.~N.~Lipatov, 
Sov.\ J.\ Nucl.\ Phys.\  {\bf 23} (1976) 338, 
%%CITATION = YAFIA,23,642;%%
%
E.~A.~Kuraev, L.~N.~Lipatov, V.~S.~Fadin,
Phys.\ Lett.\  B {\bf 60} (1975) 50, 
%%CITATION = PHLTA,B60,50;%%
Sov.\ Phys.\ JETP {\bf 44} (1976) 443,
%%CITATION = ZETFA,71,840;%%
Sov.\ Phys.\ JETP {\bf 45} (1977) 199;
%%CITATION = ZETFA,72,377;%%
%
I.~I.~Balitsky, L.~N.~Lipatov, 
Sov.\ J.\ Nucl.\ Phys.\  {\bf 28} (1978) 822. 
%%CITATION = YAFIA,28,1597;%%
%\cite{Fadin:1993qb}

\bibitem{Trajectory}
%\cite{Fadin:1996tb}
%\bibitem{Fadin:1996tb}
  V.~S.~Fadin, R.~Fiore, M.~I.~Kotsky,
%  ``Gluon Regge trajectory in the two loop approximation'', 
  Phys.\ Lett.\  {\bf B387 } (1996)  593-602.
%  [hep-ph/9605357].
%\cite{N4traj}
%\bibitem{N4traj}
  A.~V.~Kotikov, L.~N.~Lipatov,
 % ``NLO corrections to the BFKL equation in QCD and in supersymmetric gauge theories'', 
  Nucl.\ Phys.\  {\bf B582 } (2000)  19-43.
  %  [hep-ph/0004008].
 %\cite{Bartels:2008ce}
%\bibitem{Bartels:2008ce}
  J.~Bartels, L.~N.~Lipatov, A.~Sabio Vera,
%  ``BFKL Pomeron, Reggeized gluons and Bern-Dixon-Smirnov amplitudes'', 
  Phys.\ Rev.\  {\bf D80 } (2009)  045002.
%  [arXiv:0802.2065 [hep-th]].
%\cite{Bartels:2008sc}
%\bibitem{Bartels:2008sc}
%  J.~Bartels, L.~N.~Lipatov, A.~Sabio Vera,
%  ``N=4 supersymmetric Yang Mills scattering amplitudes at high energies: The Regge cut contribution'', 
  Eur.\ Phys.\ J.\  {\bf C65 } (2010)  587-605. 
%  [arXiv:0807.0894 [hep-th]].  

  \bibitem{BFKLNLO}
  V.~Fadin, L.~Lipatov, Phys.\ Lett.\  B {\bf 429} (1998) 127;
  %%CITATION = PHLTA,B429,127;%%
%
%\bibitem{Ciafaloni:1998gs}
 M.~Ciafaloni, G.~Camici, Phys.\ Lett.\  B {\bf 430} (1998) 349.
  %%CITATION = PHLTA,B430,349;%%

%\cite{Bartels:1995kf}
\bibitem{Bartels:1995kf}
  J.~Bartels, L.~N.~Lipatov, M.~Wusthoff,
  %``Conformal Invariance of the Transition Vertex $2 \to 4$ gluons,''
  Nucl.\ Phys.\  B {\bf 464} (1996) 298.
%  [arXiv:hep-ph/9509303].
  %%CITATION = NUPHA,B464,298;%%
    
%\cite{LevSeff}
%\cite{Lipatov:1995pn}
\bibitem{LevSeff}
  L.~N.~Lipatov,
  %``Gauge invariant effective action for high-energy processes in QCD,''
  Nucl.\ Phys.\  {\bf B452 } (1995)  369-400,
%  [arXiv:hep-ph/9502308 [hep-ph]].
%\cite{Lipatov:1996ts}
%\bibitem{Lipatov:1996ts}
%  L.~N.~Lipatov,
 %``Small x physics in perturbative QCD,''
  Phys.\ Rept.\  {\bf 286 } (1997)  131-198.
%  [hep-ph/9610276].
%\cite{Antonov:2004hh}
%\bibitem{Antonov:2004hh}
  E.N.Antonov, L.N.Lipatov, E.A.Kuraev, I.O.Cherednikov,
  %``Feynman rules for effective Regge action,''
  Nucl.\ Phys.\  {\bf B721 } (2005)  111-135.
%  [hep-ph/0411185].

%\cite{Martin}
\bibitem{Martin}
%\cite{Hentschinski:2008rw}
%\bibitem{Hentschinski:2008rw}
  M.~Hentschinski,
  %``Unitarity corrections from the high energy QCD effective action,''
  Acta Phys.\ Polon.\  {\bf B39 } (2008)  2567-2570;
%\cite{Hentschinski:2009zz}
%\bibitem{Hentschinski:2009zz}
%  M.~Hentschinski,
  %``The high energy behavior of QCD: The effective action and the triple-Pomeron-vertex,''
   arXiv:0908.2576 [hep-ph];
   %  [arXiv:0808.3082 [hep-ph]].
%\cite{Hentschinski:2009ga}
%\bibitem{Hentschinski:2009ga}
%  M.~Hentschinski,
  %``The effective action and the triple Pomeron vertex,''
  Nucl.\ Phys.\ Proc.\ Suppl.\  {\bf 198 } (2010)  108-111.
%  [arXiv:0910.2981 [hep-ph]].
   %\cite{Hentschinski:2008im}
%\bibitem{Hentschinski:2008im}
  M.~Hentschinski, J.~Bartels, L.~N.~Lipatov,
  %``Longitudinal loop integrals in the gauge invariant effective action for high energy QCD,'' 
  [arXiv:0809.4146 [hep-ph]]. 
     
  %\cite{Vera:2006un}
\bibitem{Vera:2006un}
  A.~Sabio Vera,
  %``The Effect of NLO conformal spins in azimuthal angle decorrelation of jet
  %pairs,''
  Nucl. Phys. B {\bf 746}, 1 (2006), 
%  [arXiv:hep-ph/0602250].
  %%CITATION = NUPHA,B746,1;%%
%\cite{Bartels:2006hg}
%\bibitem{Bartels:2006hg}
  J.~Bartels, A.~Sabio Vera, F.~Schwennsen,
  %``NLO inclusive jet production in k(T)-factorization,''
  JHEP {\bf 0611} (2006) 051. 
%  [hep-ph/0608154].
  %%CITATION = HEP-PH/0608154;%%  
%  
%\cite{Vera:2007kn}
%\bibitem{Vera:2007kn}
  A.~Sabio Vera, F.~Schwennsen,
  %``The Azimuthal decorrelation of jets widely separated in rapidity as a test
  %of the BFKL kernel,''
  Nucl. Phys. B {\bf 776}, 170 (2007).
%  [arXiv:hep-ph/0702158].
  %%CITATION = NUPHA,B776,170;%%
  %\cite{Vera:2007dr}
%\bibitem{Vera:2007dr}
  A.~Sabio Vera, F.~Schwennsen,
  %``Azimuthal decorrelation of forward jets in Deep Inelastic Scattering,''
  Phys.\ Rev.\ D {\bf 77} (2008) 014001.
%  [arXiv:0708.0549 [hep-ph]].
  %%CITATION = ARXIV:0708.0549;%%
  %\cite{Chachamis:2009ks}
%\bibitem{Chachamis:2009ks}
  G.~Chachamis, M.~Hentschinski, A.~Sabio Vera, C.~Salas,
  %``Exclusive central production of heavy quarks at the LHC,''
  arXiv:0911.2662 [hep-ph].
  %%CITATION = ARXIV:0911.2662;%%
  %\cite{Chachamis:2011rw}
%\bibitem{Chachamis:2011rw}
  G.~Chachamis, M.~Deak, A.~Sabio~Vera, P.~Stephens,
  %``A Comparative study of small x Monte Carlos with and without QCD coherence effects,''
  Nucl.\ Phys.\ B {\bf 849} (2011) 28. 
%  [arXiv:1102.1890 [hep-ph]].
  %%CITATION = ARXIV:1102.1890;%%
  %\cite{Colferai:2010wu}
%\bibitem{Colferai:2010wu}
  D.~Colferai, F.~Schwennsen, L.~Szymanowski, S.~Wallon,
  %``Mueller Navelet jets at LHC - complete NLL BFKL calculation,''
  JHEP {\bf 1012 } (2010)  026.
%  [arXiv:1002.1365 [hep-ph]].
%\cite{Angioni:2011wj}
%\bibitem{Angioni:2011wj}
  M.~Angioni, G.~Chachamis, J.~D.~Madrigal, A.~Sabio Vera,
  %``Dijet Production at Large Rapidity Separation in N=4 SYM,''
  Phys.\ Rev.\ Lett.\  {\bf 107} (2011) 191601.
%  [arXiv:1106.6172 [hep-th]].
  %%CITATION = ARXIV:1106.6172;%%%\cite{Caporale:2011cc}
%\bibitem{Caporale:2011cc}
  F.~Caporale, D.~Y.~Ivanov, B.~Murdaca, A.~Papa, A.~Perri,
  %``The next-to-leading order jet vertex for Mueller-Navelet and forward jets
  %revisited,''
  arXiv:1112.3752 [hep-ph].
  %%CITATION = ARXIV:1112.3752;%%

  %\cite{Hentschinski:2011xg}
\bibitem{Hentschinski:2011xg}
  M.~Hentschinski,
  %``Pole prescription of higher order induced vertices in Lipatov's QCD
  %effective action,''
  arXiv:1112.4509 [hep-ph].
  %%CITATION = ARXIV:1112.4509;%%


%\cite{Fadin:1993qb}
\bibitem{Fadin:1993qb}
  V.~S.~Fadin, R.~Fiore, A.~Quartarolo,
%  ``Radiative corrections to quark quark reggeon vertex in QCD",
  Phys.\ Rev.\  D {\bf 50} (1994) 2265. 
%  [arXiv:hep-ph/9310252].
  %%CITATION = PHRVA,D50,2265;%%
 
%\cite{DelDuca:1998kx}
\bibitem{DelDuca:1998kx}
  V.~Del Duca, C.~R.~Schmidt,
%  ``Virtual next-to-leading corrections to the impact factors in the high-energy limit'', 
  Phys.\ Rev.\  D {\bf 57} (1998) 4069. 
%  [arXiv:hep-ph/9711309].
  %%CITATION = PHRVA,D57,4069;%% 
  
  %\cite{Fadin:1996tb}
\bibitem{Fadin:1996tb}
  V.~S.~Fadin, R.~Fiore, M.~I.~Kotsky,
  %``Gluon Regge trajectory in the two-loop approximation,''
  Phys.\ Lett.\  B {\bf 387} (1996) 593. 
%  [arXiv:hep-ph/9605357].
  %%CITATION = PHLTA,B387,593;%%

%\cite{Laporta:2001dd}
\bibitem{Laporta:2001dd}
  S.~Laporta,
%  ``High-precision calculation of multi-loop Feynman integrals by  difference equations'', 
  Int.\ J.\ Mod.\ Phys.\  A {\bf 15} (2000) 5087. 
%  [arXiv:hep-ph/0102033].
  %%CITATION = IMPAE,A15,5087;%%

%\cite{Smirnov:2008iw}
\bibitem{Smirnov:2008iw}
  A.~V.~Smirnov,
%  ``Algorithm FIRE -- Feynman Integral REduction'', 
  JHEP {\bf 0810} (2008) 107. 
%  [arXiv:0807.3243 [hep-ph]].
  %%CITATION = JHEPA,0810,107;%%
  
%\cite{Andersen:2004uj}
\bibitem{SUSY}
  J.~R.~Andersen, A.~Sabio Vera,
  %``The Gluon Green's function in N=4 supersymmetric Yang-Mills theory,''
  Nucl.\ Phys.\ B {\bf 699} (2004) 90.
%  [hep-th/0406009].
  %%CITATION = HEP-TH/0406009;%%
%\cite{Bartels:2008ce}
%\bibitem{Bartels:2008ce}
  J.~Bartels, L.~N.~Lipatov, A.~Sabio Vera,
  %``BFKL Pomeron, Reggeized gluons and Bern-Dixon-Smirnov amplitudes,''
  Phys.\ Rev.\ D {\bf 80} (2009) 045002; 
%  [arXiv:0802.2065 [hep-th]].
  %%CITATION = ARXIV:0802.2065;%%    
    %\cite{Bartels:2008sc}
%\bibitem{Bartels:2008sc}
%  J.~Bartels, L.~N.~Lipatov, A.~Sabio Vera,
  %``N=4 supersymmetric Yang Mills scattering amplitudes at high energies: The Regge cut contribution,''
  Eur.\ Phys.\ J.\ C {\bf 65} (2010) 587.
%  [arXiv:0807.0894 [hep-th]].
  %%CITATION = ARXIV:0807.0894;%%
%\cite{Chachamis:2011nz}
%\bibitem{Chachamis:2011nz}
  G.~Chachamis, A.~Sabio Vera,
  %``The colour octet representation of the non-forward BFKL Green function,''
  arXiv:1112.4162 [hep-th].
  %%CITATION = ARXIV:1112.4162;%%
%\cite{Bartels:2009yt}
%\bibitem{Bartels:2009yt}
  J.~Bartels, C.~Ewerz, M.~Hentschinski, A.~-M.~Mischler,
  %``High Energy Behavior of a Six-Point R-Current Correlator in N = 4 Supersymmetric Yang-Mills Theory,''
  JHEP {\bf 1005} (2010) 018.
%  [arXiv:0912.4759 [hep-th]].
  %%CITATION = ARXIV:0912.4759;%%
%\cite{Bartels:2009ms}
%\bibitem{Bartels:2009ms}
  J.~Bartels, M.~Hentschinski, A.~-M.~Mischler,
  %``The Topology of the triple Pomeron vertex in N=4 SYM,''
  Phys.\ Lett.\ B {\bf 679} (2009) 460.
%  [arXiv:0906.3640 [hep-ph]].
  %%CITATION = ARXIV:0906.3640;%%  
%\cite{AlvarezGaume:2008qs}

\bibitem{gravity}
  L.~Alvarez-Gaume, C.~Gomez, A.~Sabio Vera, A.~Tavanfar, M.~A.~Vazquez-Mozo,
  %``Critical gravitational collapse: towards a holographic understanding of the Regge region,''
  Nucl.\ Phys.\ B {\bf 806} (2009) 327. 
%  [arXiv:0804.1464 [hep-th]].
  %%CITATION = ARXIV:0804.1464;%%
  %\cite{Lipatov:2011ab}
  L.~N.~Lipatov,
  %``Effective action for the Regge processes in gravity,''
  arXiv:1105.3127 [hep-th].
  %%CITATION = ARXIV:1105.3127;%%
%\cite{SabioVera:2011wy}
%\bibitem{SabioVera:2011wy}
  A.~Sabio Vera, E.~S.~Campillo, M.~A.~Vazquez-Mozo,
  %``Graviton emission in Einstein-Hilbert gravity,''
  arXiv:1112.4494 [hep-th].
  %%CITATION = ARXIV:1112.4494;%%  
  
%\cite{Romagnoni:2011jr}
\bibitem{Romagnoni:2011jr}
  A.~Romagnoni, A.~Sabio Vera,
  %``A hidden BFKL / XXX s = -1/2 spin chain mapping,''
  arXiv:1111.4553 [hep-th].
  %%CITATION = ARXIV:1111.4553;%%

\end{thebibliography}
\end{document}